\documentclass[aps,pra,twocolumn,notitlepage,superscriptaddress,tightenlines,longbibliography]{revtex4-1}
\usepackage{amsmath,amsfonts,amssymb,color,graphics,graphicx,latexsym,theorem,url,verbatim,multirow,comment}
\usepackage[]{natbib}
\usepackage[caption=false]{subfig}
\usepackage[capitalise]{cleveref}

\newcommand{\bra}[1]{\langle#1|}
\newcommand{\ket}[1]{|#1\rangle}

\newcommand{\braket}[2]{\langle#1|#2\rangle}

\begin{document}

\title{Optimizing  QAOA:  Success Probability and Runtime Dependence on Circuit Depth}

\author{Murphy Yuezhen Niu}
\email{email: murphyniu@google.com}
\affiliation{Research Laboratory of Electronics, Massachusetts Institute of Technology, Cambridge, Massachusetts, 02139, USA}
\affiliation{Department of Physics, Massachusetts Institute of Technology, Cambridge, Massachusetts 02139, USA}
\affiliation{Google Inc., 340 Main Street, Venice, CA 90291}
\author{Sirui Lu}
\affiliation{Department of Physics, Tsinghua University, Beijing, 100084, China}
\author{Isaac L. Chuang}
\affiliation{Research Laboratory of Electronics, Massachusetts Institute of Technology, Cambridge, Massachusetts, 02139, USA}
\affiliation{Department of Physics, Massachusetts Institute of Technology, Cambridge, Massachusetts 02139, USA}

\date{\today}

\begin{abstract}
 
The quantum approximate optimization algorithm~(QAOA) first proposed by Farhi et al. promises near-term applications based on its simplicity, universality, and provable optimality.  A depth-$p$ QAOA consists of $p$ interleaved unitary transformations induced by two mutually non-commuting Hamiltonians.  A long-standing   question concerning the performance of QAOA is the dependence of its success probability as a function of circuit depth $p$. We make initial progress   by analyzing the success probability of QAOA for realizing state transfer in a one-dimensional qubit chain   using two-qubit XY Hamiltonians and single-qubit Hamiltonians. We provide analytic state transfer success probability dependencies on $p$ in both low and large $p$ limits by leveraging the unique spectral property of the XY Hamiltonian. We support our proof under a given QAOA ansatz with numerical optimizations of  QAOA for up to \(N\)=20 qubits. We show that the optimized QAOA can achieve the well-known quadratic speedup, Grover speedup, over the classical alternatives. Treating the QAOA optimization as a quantum control problem, we also provide numerical evidence of how the circuit depth determines the controllability of the QAOA ansatz.

%We also study the circuit depth scaling, and the scaling with physical run time of QAOA in this simple model. 

\end{abstract}
\maketitle
%\tableofcontents

\section{Introduction}\label{introduction}

%structure of introduction:
%
%1. why QAOA is important
%2. what're the major open questions in QAOA
%3. how are we addressing these questions with the example of state transfer

%
%Quantum adiabatic theorem~\citep{PhysRevE.58.5355} states that if the time-dependent Hamiltonian changes sufficiently slow, the initial state prepared in the ground state remains in the ground state throughout the evolution. This principle offers a shortcut to solving NP-complete problems by encoding the solution to combinatorial optimization problems in the ground state of the end Hamiltonian, which is known as quantum adiabatic algorithm (QADI)~\citep{farhi2000quantum, farhi2001quantum}. However, the primary drawback of this method is the possible long run-time required to maintain the instantaneous ground state. It is also hard to analyze the run-time of a specific problem.

Quantum approximate optimization algorithm~(QAOA) promises near-term applications for quantum devices given its simplicity, universality and provable optimality.
In contrast to quantum adiabatic algorithms~\citep{farhi2000quantum, farhi2001quantum}, QAOA adopts abrupt switching between two different Hamiltonian evolutions~\citep{farhi2014quantum}. This simple strategy can reduce the complexity of Hamiltonian controls by obviating the need to continuously vary Hamiltonian in time. Fine-tuned controls over Hamiltonian trajectories  are otherwise  necessary   for the   traditional quantum adiabatic algorithms. Despite its simplicity,
QAOA is computationally universal~\citep{lloyd2018quantum}.
 It also has important implications in computational complexity: the efficient classical sampling of a depth-1 QAOA will collapse the polynomial hierarchy to the third level~\citep{farhi2016quantum}.   Lastly, from Pontryagin's maximum principle~\citep{stengel2012optimal}, QAOA is optimal for solving variational problems whose cost function is a linear function of the system Hamiltonian~\citep{yang2017optimizing}. 

%
%The Pontryagin's maximum principle states that if the control Hamiltonian is a linear function of control parameters, the optimal control solution is of bang-bang form, equivalent to switching the Hamiltonian governing the state evolution in a discrete way same as that used in QAOA.

Despite these attractive properties,  to be suitable for near-term quantum devices, a long-standing  question remains to be addressed: how does the success probability of QAOA depend on its circuit depth? Near-term quantum device's computation time is limited by noise and decoherence.   This in turn limits the realizable  quantum algorithms to relatively short circuit depth.   To  understand QAOA's   potential or limitations for near-term applications, it is therefore critical to understand its performance when fixing the upper bound on the depth of the QAOA circuit.  

It is exceedingly hard to study the QAOA performance without choosing the QAOA Hamiltonian and optimization problem. This is mainly due to the lack of a sufficient condition  for QAOA to achieve optimality in a generic scenario. Since the QAOA success probability directly depends on its optimality, a bound on QAOA success probability scaling usually requires problem-specific numerical optimizations. Recently,   specific properties of the chosen optimization problem and QAOA Hamiltonians are utilizied to design protocols that imitate the Grover search algorithm~\citep{PhysRevA.95.062317,smelyanskiy2018non}, or to prepare highly entangled quantum states~\citep{ho2018efficient,ho2018ultrafast}. These encouraging results spotlight the importance of  the problem and hardware specific information such as the controllable system Hamiltonians in designing QAOA algorithms for improving its performance guarantee.

%Besides protocol based on quantum optical systems, one can also use condensed matter system like qubit chain~\citep{bose2003quantum,PhysRevLett.106.040505}. There are many proposals using quantum wire so far like disorder system~\citep{PhysRevLett.99.167201, PhysRevA.80.052319}, optimal control~\citep{zhang2016optimal}, long-range interaction system~\citep{richerme2014non} and adiabatic evolution~\citep{PhysRevA.77.012303} under moving potential.

In this work, we make initial progress towards answering this  question by   analyzing the performance of QAOA for   state transfer in a one-dimensional qubit chain using the XY Hamiltonians and single-qubit Hamiltonians as QAOA ansatzes. 
We choose quantum state transfer as our QAOA optimization task considering its simplicity and wide applications~\citep{bose2003quantum,PhysRevLett.106.040505,PhysRevLett.99.167201, PhysRevA.80.052319,richerme2014non,PhysRevA.77.012303,christandl2005perfect}. State transfer is  a preliminary requirement for realizing quantum networks which are necessary for connecting quantum computers to form large-scale computation network~\cite{PhysRevLett.78.3221,kimble2008quantum}.  We choose the QAOA Hamiltonian ansatz based on the experimental availability, where   XY couplings are available in existing superconducting qubit device~\cite{PhysRevLett.112.200501}. Moreover, the XY Hamiltonian's particle number conserving nature makes it suitable for realizing  state transfer within a given particle number subspace, which can mitigate unwanted information leakage into the higher excitation subspace. 

We harness the analytic spectral features of the XY Hamiltonians to derive explicit success probability $P_{\text{succ}}$ dependencies on circuit depth $p$ for   state transfer in two different limits.  In the  low circuit depth and short QAOA duration limits we have  $\lim_{p\to 0, \delta\to 0}P_{\text{succ}}\propto p^{2}$; and in the large circuit depth limit we have $\lim_{p\to \infty, \delta\to 0}P_{\text{succ}}\propto p^{4p}$.  Our proof reconfirms the achievable Grover speedup with QAOA ansatz~\cite{PhysRevA.95.062317}. As a compliment to the system-specific analysis, we apply the existing results of the Lieb-Robinson bound to the QAOA success probability with any 2-local bounded-norm qubit Hamiltonians for one-dimensional state transfer.  
To verify the optimality of our scaling proof, we numerically optimize the associated QAOA ansatz for different circuit depth and overall runtime. Our numerical results confirm the expected quadratic Grover-like scaling. We  also demonstrate an interesting connection between the achievable success probability and its controllability dependency on the circuit depth: once the circuit depth is too low, the QAOA is no longer controllable when the control landscape is full of local optima that are not globally optimal. 

The structure of the paper is as follows: in Sec.~\ref{qaoa-for-state-transfer} we introduce the basic setup of the QAOA for state transfer using the XY Hamiltonian;  in Sec.~\ref{SuccessScalingSection}  we derive the QAOA success probability as a function of circuit depth  in both low  and large circuit depth limit;  we discuss the associated quantum speed limit   in Sec.~\ref{quantum-speed-limit}; we summarize the performance of the QAOA in regard to circuit depth, runtime and number of qubits   in Sec.~\ref{NumericalSection}, and conclude in Sec.~\ref{SummarySection}.

%
%Then we address these following question: `Is optimal QAOA capable of achieving quantum speed limit?' and `How powerful is low depth QAOA circuit?'. To address these questions, we have performed numerical simulation over different circuit depth, physical run time and the number of qubits.

\nocite{zhou2018quantum,pichler2018computational,pichler2018quantum,brandao2018fixed}

\section{QAOA for State Transfer}\label{qaoa-for-state-transfer}

We introduce in this section the basic concept of quantum state transfer and its realization through QAOA. Quantum state transfer has been proposed in both quantum optical systems and condensed matter system~\citep{bose2003quantum,PhysRevLett.106.040505}. Different state transfer schemes include the use of quantum disorder~\citep{PhysRevLett.99.167201, PhysRevA.80.052319}, optimal control~\citep{zhang2016optimal}, long-range interaction~\citep{richerme2014non} and adiabatic evolution under a moving potential~\citep{PhysRevA.77.012303}. Unlike the majority of these existing approaches, QAOA resorts to a discrete set of operations of a size given by the QAOA circuit depth. Such a circuit-based model is naturally suitable for near-term quantum devices such as superconducting qubits and ion traps, but more flexible than the traditional circuit-based model using gates only from a predefined universal gate set.  
\nocite{li2018perfect} % a state transfer experiment using the superconducting qubit platform. 

The state transfer problem of interest  is defined in a one-dimensional qubit chain of length \(N\). We  use \(|\overline{n}\rangle\) to represent a product state of a positive eigenstate of the local Pauli-\(z\) operator at the \(n\)th site and the negative eigenstates of the local Pauli-\(z\) of other sites: \(\sigma_{n}^z|\overline{n}\rangle=|\overline{n}\rangle,~\sigma^z_{i,i\neq n}|\overline{n}\rangle=-|\overline{n}\rangle\), e.g. $|\overline{n}\rangle = |0\rangle_1 |0\rangle_2 \cdots |0\rangle_{n-1} |1\rangle_{n} |0\rangle_{n+1} \cdots |0\rangle_N$. We denote  \(|\overline{0}\rangle\) as the product state of the negative eigenvalue eigenstate of the local Pauli-\(z\) operators: \(|\overline{0}\rangle=|0\rangle_1|0\rangle_2\cdots |0\rangle_N\). If we treat qubits as spins, and negative eigenvalue of Pauli-\(z\) operator as an excitation of the spin state, the state transfer problem we will solve lies in the span of zero and single excitation subspaces. In this subspace, a  quantum state with boundary excitation is represented as
\begin{align}
\ket{\psi_i}= \alpha \ket{\overline{1}} + \beta \ket{\overline{0}},
\end{align}
with $\vert \alpha \vert^2 + \vert\beta\vert^2 =1$. 
Starting from the  state $|\psi_i\rangle$, the task of state transfer is therefore to realize a unitary transformation \(U\) such that:
\begin{align}
\ket{\psi_f}=U\ket{\psi_i}=\alpha \ket{\overline{N}} + \beta \ket{\overline{0}}.
\end{align}

We choose the two Hamiltonians used for QAOA iteration to be
\begin{align}
\hat{H}_C  = |\overline{N}\rangle\langle \overline{N} | = \frac{1}{2}(\sigma_N^z+I_N),\\
\hat{H}_B = \sum_{i=1}^N (\sigma_i^x\sigma_{i+1}^x+\sigma_{i}^y\sigma_{i+1}^y).
\label{eq:hcb}
\end{align}
The reasons for our choice of QAOA Hamiltonians are two-fold: (1) $\hat{H}_C$'s  plus one eigenstate is our transferred target state $\ket{\overline{N}}$ and can thus serve as a Grover-like oracle by assigning a phase to the target state; (2) $\hat{H}_B$ is off-diagonal and induces a swap operation between neighboring qubits, and thus can move the excitation around for the purpose of state transfer.

Since the total qubit-\(z\) operator \(S_z=\sum_{i=1}^N \sigma_i^z\) commutes with both \(\hat{H}_C\) and \(\hat{H}_B\) and $|\overline{1}\rangle$ is an eigenstate of the total qubit-\(z\) operator: \(S_z |\overline{1}\rangle = (1-N)|\overline{1}\rangle\), the total excitation is conserved throughout the QAOA simulation. We can therefore solve the quantum dynamics in the subspace spanned by \(\{|\overline{0}\rangle,|\overline{1}\rangle,|\overline{2}\rangle, \cdots,|\overline{N}\rangle\}\).

Denoting the unitary evolution under \(\hat{H}_C\) for a time duration \(t\) as \(U(\hat{H}_C,t)\) and the unitary evolution under \(\hat{H}_B\) for time duration \(t'\) as \(U(\hat{H}_B,t')\), a depth \(p\) QAOA  realizes the following unitary transformation:
\begin{align}
U_p = \prod_{k=1}^p U(\hat{H}_C, \delta_k^C) U(\hat{H}_B, \delta_k^B),
\end{align} 
where the durations of evolutions under given Hamiltonians are represented by $\delta_k^{B}$ and $\delta_k^{C}$. Here, each $k$th QAOA iteration consists of a unitary evolution under \(\hat{H}_B\) for time \(\delta_k^B \) then followed by a unitary evolution under  \(\hat{H}_C\) for time \(\delta_k^C \). Since the eigenvalues of $\hat{H}_B$ are not rational, unlike  the original QAOA, the  rotation angle  $\delta_k^B$ of $\hat{H}_B$ is  not restricted to $(0,2\pi)$.

Since the zero excitation state of the system is invariant under the unitary evolution of both Hamiltonians, the state transfer task is equivalent to realizing the unitary transformation: $\ket{\overline{1}} \to \ket{\overline{N}}$. Thus, we can quantify the fidelity of state transfer by the fidelity between \(U_p|\overline{1}\rangle\) and \(|\overline{N}\rangle\):
\begin{equation}
F = |\bra{\overline{N}}U_p |\overline{1}\rangle|^2=\langle\overline{1}|U_p^\dagger \hat{H}_CU_p|\overline{1}\rangle .
\end{equation}
This is equivalent to the success probability, so we will use them interchangeably henceforth. We can then treat state transfer as a special kind of maximization satisfaction problem, except that the cost function \(F\) here contains only one clause as opposed to many clauses in traditional QAOAs~\cite{farhi2014quantum}.  

%The fidelity (i.e., success probability) of realizing the state transfer with \(p\) switches is

It is shown in~\citep{christandl2005perfect} that if we have complete control over the amplitudes of XY coupling at different sites, perfect state transfer can be realized through a single unitary evolution under the XY Hamiltonian. This is, however, unrealistic for near-term devices, where the system calibration for   such fine-tuned interactions is costly and the maximum interaction strength is limited. 
% \(p-1\)-repetitions of \(U(\hat{H}_B,t_i')U(\hat{H}_C,t_i)\) such that the final state writes 
%\begin{equation}
%|\psi_f\rangle = U(\hat{H}_B,t_{p}') U(\hat{H}_C,t_p)\cdots U(\hat{H}_B,t_2') U(\hat{H}_C,t_2)U(\hat{H}_B,t_1)|\overline{1}\rangle.
%\end{equation} 

%{\color{red} revise}This paper is organized as follow, in the next section, we gave a proof of the fundamental limit of state transfer for norm bounded nearest-neighbor interaction by using the Lieb-Robinson bound. The required time scales with the system size. Then we solve the dynamics of our system in the subspace spanned by \(\{|\overline{1}\rangle,|\overline{2}\rangle, \cdots,|\overline{n}\rangle\}\). Next, we fix the physical run time and numerically study the circuit depth scaling of success probability.

\section{Success Probability Scaling as a Function of Circuit Depth}\label{SuccessScalingSection}
In this section, we derive the success probability scaling as a function of circuit depth.   Our analysis is based on an iterative procedure using the knowledge from the spectrum of the QAOA ansatz Hamiltonian.

%We define the depth of QAOA as the number of applications of the oracle and dispersion Hamiltonians, such that a depth $p$ QAOA results in a unitary of the form:
%\begin{align}
%U_p = \prod_{k=1}^p U(C, \delta_k^C) U(B, \delta_k^B)
%\end{align} where the duration of evolution under a given Hamiltonian is represented by $\delta_k^{B/C}$.  This definition of depth is equivalent to circuit depth up to a factor of two which will be used interchangeably.   

To simplify our analysis, we adopt the  following QAOA ansatz: the duration under the evolution of dispersion Hamiltonian $\hat{H}_B$, $\delta$, is short and the same for different iterations, and the evolution under the diagonal Hamiltonian $\hat{H}_C$ is of angle $\pi$, resembling a Grover oracle: $U_p =  \left( e^{-i\pi \ket{\overline{N}}\bra{\overline{N}}} U(\hat{H}_B, \delta ) \right)^p$.  With this ansatz, our result can be connected to the scaling analysis in conventional Grover search and thus serves as a lower bound on the success probability for the optimized QAOA to be discussed in the subsequent sections. 

We first diagonalize the dispersion Hamiltonian $\hat{H}_B$ in the single excitation subspace to obtain its $k$th eigenstate:
\begin{widetext}
\begin{equation}\label{eq:eigenstate_HB}
\ket{\phi_k}=\frac{1}{\sqrt{N/2}}\sum_{n=1}^{N/2}\left( \sin\left[\frac{kn\pi}{N/4+1} \right] \ket{\overline{2n}} +  \sin\left[\frac{k(n + 1/2)\pi}{N/4+1} \right]\ket{\overline{2n-1}} \right),
\end{equation}
\end{widetext}
with the $k$th eigenvalue being $E_k = 2 \cos[\frac{k\pi}{N/2 + 1}]$. 

Given the initial state of the system as $\ket{\psi_0}= \ket{\overline{1}}$, the system evolves to $\ket{\psi_1}=e^{-i\hat{H}_B\delta}\ket{\overline{1}}$ after a depth one QAOA. Then the success probability of transferring the excitation to the other end of chain after a depth one QAOA is
\begin{equation}
P_{\text{succ}}(1)=\langle \overline{1} |e^{i\hat{H}_B\delta}|\overline{N} \rangle\langle \overline{N}|e^{-i\hat{H}_B\delta}\ket{\overline{1}}=|f_1^N(\delta)|^2,
\end{equation}
where we use $ f_1^N(\delta)=\bra{\overline{N}}e^{-i\hat{H}_B\delta}\ket{\overline{1}} $ to represent the amplitude of target state. Now we apply another QAOA iteration to update the quantum system to
\begin{equation}
\begin{split}
\ket{\psi_2} &=U(\hat{H}_B,\delta) U(\hat{H}_C, \pi)\ket{\psi_1}\\
&=e^{-i2\hat{H}_B\delta}\ket{\overline{1}} - 2e^{-i\hat{H}_B\delta}\ket{\overline{N}}\bra{\overline{N}}e^{-i\hat{H}_B\delta}\ket{\overline{1}}
\end{split}
\end{equation}
This gives  the success  probability of the  state transfer after a depth two QAOA as:
\begin{equation}
P_{\text{succ}}(2)= \langle \psi_2 |\overline{N}\rangle\langle \overline{N}|\psi_2\rangle =|f_1^N(2\delta) -2f_1^N(\delta)f_N^N(\delta)|^2,
\end{equation}
where $ f_N^N(\delta)= \bra{\overline{N}}e^{-i\hat{H}_B\delta}\ket{\overline{N}} $ denotes the amplitude of the state $ \ket{\overline{N}} $ remaining in state $ \ket{\overline{N}} $ after Hamiltonian evolution under $ \hat{H}_B $ for time $ \delta $. Similarly, we continue the iteration to obtain the success probability after a depth three and depth four QAOA:
\begin{widetext}
\begin{align}
P_{\text{succ}}(3)&=f_1^N(3\delta)-2f_1^N(2\delta)f_N^N(\delta) - 2f_1^N(\delta)f_N^N(2\delta) + 4 \left(f_N^N(\delta)\right)^2f_1^N(\delta),\\
\begin{split}
    P_{\text{succ}}(4)&=f_1^N(4\delta)-2f_1^N(3\delta)f_N^N(\delta) - 2f_1^N(2\delta)f_N^N(2\delta)- 2f_1^N(\delta)f_N^N(3\delta) + 4\left(f_N^N(\delta)\right)^2f_1^N(2\delta)
\\
    &+4\left(f_N^N(\delta) \right)^2 f_N^N(2\delta)  + 4f_N^N(2\delta) f_N^N(\delta)f_1^N(\delta)-8\left(f_N^N(\delta)\right)^3f_1^N(\delta).
\end{split}
\end{align}
\end{widetext}
Now we provide expression for transition amplitude used above given the exact eigenstates of the dispersion Hamiltonian $\hat{H}_B$ in Eq.~\ref{eq:eigenstate_HB}:
\begin{widetext}
\begin{align}
\begin{split}
    f_1^N(\delta)&=\sum_{k=1}^{N/2}\braket{\overline{N}}{\phi_k}\braket{\phi_k}{\overline{1}}e^{-i E_k \delta} ,\\
   &\approx-\frac{i \delta}{2\sqrt{N+1}}  \left(\cos \left(\frac{2 \pi  (N/2)^2+4 \pi  N/2+\pi }{2 (N/4+1)}\right)
   \csc \left(\frac{\pi  N/2+2 \pi }{2 (N/4+1)}\right)+\cos \left(\frac{\pi }{2
   (N/4+1)}\right) \csc \left(\frac{\pi  N/2+2 \pi }{2 (N/4+1)}\right)\right)\\
   &= -iF(N)\delta,
\end{split}\\
\begin{split}
   f_N^N(\delta)=&\sum_{k=1}^{N/2}\braket{\overline{N}}{\phi_k}\braket{\phi_k}{\overline{N}}e^{-i E_k \delta} ,\\
   \approx&\frac{i \delta}{2\sqrt{N+1}}\bigg\{ \frac{1}{2} \left[-\cos \left(\frac{4 \pi  N^2+\pi  N-\pi }{2 (N+1)}\right) \csc\left(\frac{\pi  N}{N+1}\right)+2 N+\cos \left(\frac{\pi  (N-1)}{2 (N+1)}\right) \csc\left(\frac{\pi  N}{N+1}\right)\right]\\
   &\times   \left[\cos \left(\frac{\pi  N}{2 (N+1)}\right) \csc \left(\frac{\pi }{2(N+1)}\right)+\cos \left(\frac{\pi  (N+2)}{2 (N+1)}\right) \csc \left(\frac{\pi }{2(N+1)}\right)\right]\\
   &-  \left[\cos \left(\frac{3 \pi  N}{2 (N+1)}\right) \csc\left(\frac{\pi -2 \pi  }{2 (N+1)}\right)-\cos \left(\frac{-4 \pi  N^2-\pi  N}{2(N+1)}\right) \csc \left(\frac{\pi -2 \pi  N}{2 (N+1)}\right)\right]\\
   &-  \left[\cos \left(\frac{\pi  N}{2 (N+1)}\right) \csc \left(\frac{2\pi  N+\pi }{2 (N+1)}\right)-\cos \left(\frac{4 \pi  N^2+3 \pi  N}{2 (N+1)}\right)\csc \left(\frac{2 \pi  N+\pi }{2 (N+1)}\right)\right] \bigg\}\\
   =&-iG(N) \delta
\end{split}
\end{align}
\end{widetext}
where the approximation is made to include only the terms that are of either zero or first order in $ \delta $, under the   limit $ \delta \to 0 $. And both $ F(N) $ and $ G(N) $ are  real-valued. For a QAOA of depth $ p $, we deduce the success probability  dependency on transition probabilities $f_1^N(\delta)$ and $f_N^N(\delta)$ as
\begin{equation}
    P_{\text{succ}}(p) =\sum_{j=1}^p(-1)^{j}\sum_{\vec{v}_j \in \mathcal{V}_j}f_1^N(\vec{v}_j(1)\delta )\prod_{k=1}^{j-1}f_N^N(\vec{v}_j(k+1)\delta )
\end{equation} where $\vec{v}_j$ is a vector with each element representing the value of a $j$ partition of $p$ that belongs to the set $\mathcal{V}_j=\{\vec{v}_j|\sum_{k=1}^j\vec{v}_j(k)=p\}$. The success probability can be in turn  be expressed as
\begin{equation}
   P_{\text{succ}}(p)\approx\left|\sum_{j=1}^{p+1} A_j \delta^j \right|^2, 
\end{equation}
with the each amplitude $A_j$ given by
\begin{align}
A_1&= -ipF(N),\\
A_2&=- F(N)G(N)\frac{p(p+1)(p+2)}{3},\\\label{EqasymptoticAn}
\lim_{n\to \infty} A_n&\approx -F(N)G(N)^n p^{2n-1}.
\end{align}
Eq.~(\ref{EqasymptoticAn}) is derived from the asymptotic value of the product of all possible values of $n$ integers $p_1, p_2, \ldots, p_n$ whose sum equals $p$: $\sum_{i=1}^n p_i=p$.   

So far we have only kept the leading order in $O(\delta)$ together with  all orders of $O\left(\left(p^2\delta\right)^n\right)$ which will be non-negligible when $p^2\delta\sim 1$ or $p^2\delta \gg 1$.
For the scaling analysis, we neglect constant terms in the sum and find the success amplitude to be
\begin{widetext}
\begin{equation}
\begin{split}
   \sum_{j=1}^{p+1} A_j \delta^j &= -i(p+1)F(N)\delta - \left[  F(N)G(N)p^3\delta^2 + \cdots +  F(N)G(N)^p p^{2p+1}\delta^{p+1}\right]\\
    &=-i(p+1)F(N)\delta -\frac{F(N)G(N)p^3 \delta^2\left[ (G(N)p^2\delta )^p-1\right]}{(G(N)p^2\delta )^2 -1}.
\end{split}
\end{equation}
\end{widetext}

Since the amplitude is composed of imaginary and real parts, the success probability of a depth-$p$ QAOA is thus found to be
\begin{widetext}
\begin{equation}
P_{\text{succ}}(p)\approx(p+1)^2F(N)^2\delta ^2 + \frac{F(N)^2G(N)^2p^6 \delta^4\left[ (G(N)p^2\delta )^p-1\right]^2}{\left[(G(N)p^2\delta )^2 -1\right]^2}.
\end{equation}
\end{widetext}
This success probability dependence can be used as a lower bound on the QAOA performance after optimization where the duration of each evolution can be of flexible value.
In the large depth limit, the term with the largest power of $ p $ dominates:
\begin{equation}
\lim_{p\to \infty} P_{\text{succ}}(p) \propto p^{4p+2}.
\end{equation}
This exponential growth in success probability is based on the assumption that $ \delta $ is a small constant~(which does not change with the circuit depth $p$), and the dominant contribution to the transition amplitudes is of the lowest order in $\delta$. Such exponential dependence is also found by increasing the speed of   adiabatic Hamiltonian evolution, see~\cite{Lidar2010}. We also observe such exponential growth in our numerically optimized QAOA (see Sec. \ref{SuccessScalingSection}).

In the low-depth limit, only the lowest order of $\delta$ terms dominates:
\begin{equation}
\lim_{p\to 1} P_{\text{succ}}(p) \propto F(N)^2 \delta   p^2.
\label{Pp-low-depth}
\end{equation}
Then the total number of steps required to achieve the target state is of order $ O(1/(\delta F(N)))=O(\sqrt{N}).$ This quadratic Grover-like dependence on circuit depth can be understood by mapping a Grover algorithm  to a QAOA routine.  The dispersion step of Grover iteration, which is a rotation of angle $\pi$ around the equal superposition state:
\begin{equation}
U_s=H^{\otimes N} (-2\ket{0}\bra{0} + I) H^{\otimes N},
\end{equation} with $H$ representing the Hadamard gate, 
can be generalized to a rotation around any state that is not parallel to the target state $\ket{\psi}$~\cite{nielsen2002quantum}:
\begin{equation}
U_s= -2 \ket{\psi}\bra{\psi} + I.
\end{equation}
If we choose $\ket{\psi}= e^{-i\delta \hat{H}_B}\ket{\overline{1}}$, the corresponding $p$th Grover iteration for searching the transferred state $\ket{\overline{N}}$ starting from the initial state $\ket{\overline{1}}$ can then be represented by the unitary realized by a depth $p$ QAOA circuit as 
\begin{equation}
    U_{\text{Grover}}^p = \left(e^{-i \hat{H}^1_c\pi}e^{i\hat{H}_B\delta_1}   e^{-i \hat{H}^2_c\pi}e^{-i \hat{H}_B\delta_1}\right)^p,
\end{equation}
where $\hat{H}^1_c=\frac{1}{2}(\sigma_N^z + I_N)$ and $\hat{H}^2_c=\frac{1}{2}(\sigma_1^z + I_1)$.

%, \,\text{with}\,\, \delta t_1 \ll 1.

% We numerically optimize the success probability over $\delta t_1$ using BFGS algorithm for different circuit depth $p$. Among all optimal solution for different circuit depth, we pick the smallest $p$ with $F>0.95$. The optimized circuit depth scalings for $N=3$ and $N=10$ are depicted in \cref{fit-grover7}. The optimized Grover circuit depth scaling at the beginning is polynomial.   
% %
% %%We numerically simulated this process for up to $n=100$ with system size $N=1\cdots 20$. 
% %
% \begin{figure}[h]
% \subfloat[$N=3$ qubits]{\includegraphics[width=0.48\linewidth]{figs/grover-new/95L3_max_p_60_Niter200.pdf}} 
% \subfloat[$N=10$ qubits]{\includegraphics[width=0.48\linewidth]{figs/grover-new/95L10_max_p_200_Niter200.pdf}}
% \caption{\textbf{Fidelity $F$ versus circuit depth $p$ with no constraints on $t_f$ with optimized grover-type algorithm.} We choose the first circuit depth that has optimal fidelity larger than 0.95} 
% \label{fit-grover7}
% \end{figure}

\section{Quantum Speed Limit}\label{quantum-speed-limit}
As a supplement to the success probability scaling analysis presented in the previous section that is limited to our specific choices of the QAOA Hamiltonians, we review in this section the general constraints on the QAOA performance using spatially local Hamiltonians imposed by the Lieb-Robinson bound. The Lieb-Robinson bound~\citep{lieb1972finite} is a powerful tool to study the propagation of quantum correlation and thus quantum information in many-body quantum systems~\citep{hastings2010locality}.   %The bound we derived here is applicable for a depth one QAOA without switching between the two Hamiltonians.
It   serves as a lower bound on the success probability for the QAOA performance of the same total runtime. Although such a bound is not tight, nor does it directly depends on the circuit depth of QAOA, it provides a basic reference of the optimality of QAOA in regard to its success probability scaling as a function of physical time. And in fact, the theoretical insights of Lieb-Robinson bound help us to understand the performance of numerically optimized QAOAs in the next section.

We rewrite our QAOA iterations as an evolution under the time-dependent Schr\'{o}dinger equation with the time-dependent Hamiltonian:  
\begin{equation}
\hat{H}(t)=s(t)\hat{H}_C+[1-s(t)]\hat{H}_B,
\end{equation} 
where \(s(t)\) is the time varying control parameter that can only take on the values zero and one. Thus it realizes the bang-bang form of QAOA iterations. 

%Since the interactions between different qubits in this system are two local

\begin{figure}[h]
\includegraphics[width=0.95\linewidth]{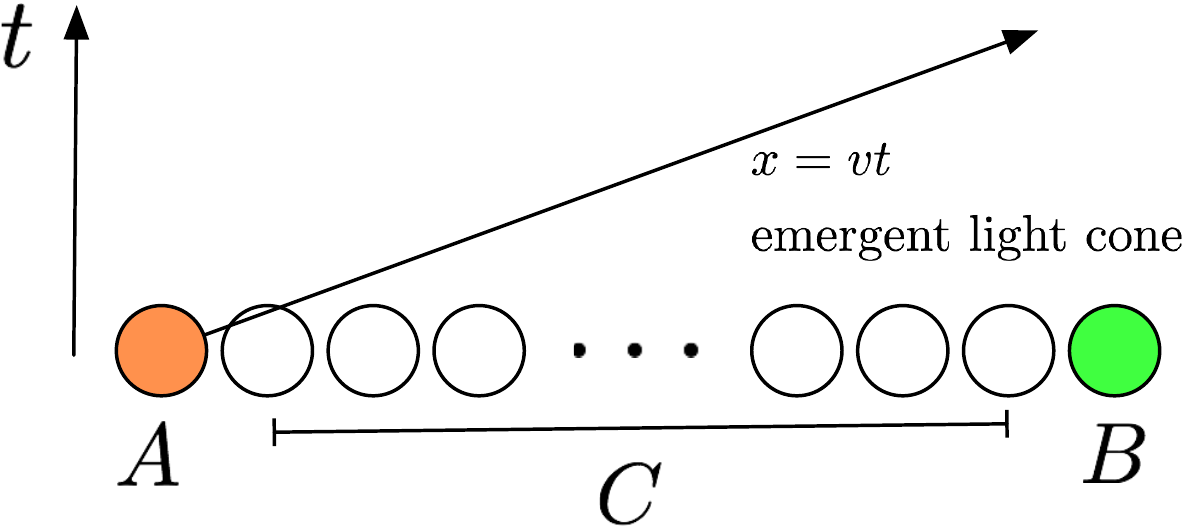}~
%\subfloat[]{\includegraphics[width=0.95\linewidth]{figs/LR.pdf}}
\caption{The emergent light cone in  state transfer problem, where  the quantum state localized at $A$ (the first qubit) is transfered to $B$ (the last qubit) through the quantum channel $C$ consisting of qubits in the middle. In the short-range two-local system, a non-relativistic light-cone $x=vt$ emerges. The amount of information that can be transferred outside of the lightcone is exponentially small.}
\label{state-transfer}
\end{figure}

Note $\hat{H}(t)$ can be written as the sum of nearest-neighbor interacion terms: $\hat{H}(t)=\sum_i h_{i,i+1}(t)$. Thus, by Lieb-Robinson bound, the maximum speed of quantum information propagation in this system is bounded. This speed determines how fast operations on the first qubit can affect observables on the last qubit at some later moment of time, and thus upper bounds the speed of state transfer. For convenience, let us call the first qubit \(A\), the last qubit \(B\), and the rest part \(C\) (see \cref{state-transfer}).  The  Lieb-Robinson  bound determines the maximum  operator norm of the commutator between any operators \(O_A\) and \(O_B\) on the first and the last qubit   at a later time $t$. More specifically, denoting \(L=N-1\) as the distance between the first and the last qubit and \(J=\max_i\max_t\|h_{i,i+1}(t)\|\) as the maximum interaction strength, the Lieb-Robinson bound for a nearest-neighbour Hamiltonian on a $D$-dimensional square lattice \cite{richerme2014non,hastings2010locality} is given by
\begin{equation}
\|[O_A(t),O_B(0)]\|\leq 2\|O_A\|\|O_B\|\sum_{k=L}^{\infty}\frac{(2Jt(4D-1))^k}{k!}.
\end{equation}
In the one-dimensional system, the above bound simplifies to \cite{richerme2014non}:
\begin{align}
\|[O_A(t),O_B(0)]\|&\leq
2\|O_A\|\|O_B\|\sum_{k=L}^{\infty}\frac{(6Jt)^k}{k!}\\
&\leq  2\|O_A\|\|O_B\|\exp({6eJt-L})\\
&= 2\|O_A\|\|O_B\|\exp{\left(vt-L\right)},
\end{align}
where the Lieb-Robinson velocity $v=6eJ$ is approximately 32.616 because $J=\|\sigma_i^x\sigma_i^x+\sigma_i^z\sigma_i^z\|=2$. Consequently, $1/v\approx 0.03$. %If we consider the group velocity of free fermion, $v=2J$, which will gives $1/v\approx 0.25$.

%, while \(c,~v\), and \(\xi\) are positive constants depending only upon   More specifically, 

%\begin{equation}
%\|[O_A(t),O_B(0)]\|\leq c %\|O_A\|\|O_B\|\exp{\left(-\frac{L-v|t|}{\xi}\right)},
%\end{equation}

%In our case, we choose our Hamiltonian to satisfy \([\hat{H}(t),\sum_i^N \sigma_i^z]=0\). 
Let \(U_A^0 = I_A\) and \(U_A^1 =\sigma _A^x\). The unitary transformation of the whole system is induced by the time-dependent Hamiltonian as
\(U_{ABC}(t)=\mathcal{T}\exp (\int_0^t -i \hat{H}(t') \mathrm{d}t')\). Specifically, with initial  state of the system given by \(\rho_0=|\overline{0}\rangle\langle \overline{0}|\), we can interpret this procedure as a quantum channel where the input state is \(\rho_{ABC}^k = U_A^k \rho_0 U_A^{k\dagger}\), and the output state, the reduced density matrix of \(B\), is \(\sigma_B^k (t) = \text{Tr}_{AC}(U_{ABC}(t)\rho_{ABC}^k U_{ABC}^\dagger(t))\).

Ref. \cite{PhysRevLett.97.050401} shows that \(\sigma_B^k(t)\) depend on \(k\) as follows. For any observable \(O_B\) and associated time evolved operator \(O_B(t)=U_{ABC}^\dagger (t)O_B U_{ABC}(t)\), we have:
\begin{equation}
\text{Tr}_B\left[O_B(\sigma_B^0(t)-\sigma_B^1(t))\right] \leq \epsilon\|O_B(t)\|,
\end{equation}
where \(\epsilon = 2 \exp(vt-L)\) is given by the Lieb-Robinson bound. Here, we have used \(U_A^{1\dagger} U_A^1=I\), the definition of operator norm, and the Lieb-Robinson bound.

Therefore \(\sigma_B^1(t)\) and \(\sigma_B^0(t)\) are \(\epsilon\) close in the trace norm: \(\|\sigma_B^1(t)-\sigma_B^0(t)\|_1\leq \epsilon\). By using the following inequality between the trace distance and the fidelity, \(F(\rho,\sigma)\geq 1-\frac{1}{2}\|\rho-\sigma\|_1\), we obtain a bound on the fidelity between \(\sigma_B^1(t)\) and \(\sigma_B^0(t)\): 
\begin{equation}
F(\sigma_B^1(t),\sigma_B^0(t))\geq 1-\frac{1}{2}\epsilon.
\end{equation} 
Now we would like to bound the success probability \(P(t)\) of a excitation of qubit state transferred from site $1$ to the site \(N\)  after the evolution under \(\hat{H}(t)\) for time \(t\) which depends on the fidelity defined above as: 
\begin{equation}
1-P(t) = F^2(\sigma_B^1(t),\sigma_B^0(t)) 
  \geq (1-\frac{1}{2}\epsilon)^2.
\end{equation} 
Rearranging, we obtain an upper bound for the success probability of the QAOA as a function of time and the length of the qubit chain:
\begin{equation}
\label{PtLRB}
P(t) \leq\epsilon-\frac{1}{4}\epsilon^2,
\end{equation} 
From the above expression, we can identify three different regions of temporal dynamics. At the early time when $t\ll L/v$, we have $\epsilon= 2 \exp(vt-L))\ll 1$ and the probability of success is nearly zero. In this first region, the success probability is   \textit{exponentially suppressed} and remains almost zero in time. When $t \approx L/v$ we have \(\epsilon = c \exp(vt-L)<1\) and the first term of the right-hand side of Eq.~(\ref{PtLRB}) dominates, which gives rise to an \textit{exponentially growing} success probability. Finally when  $t> L/v$, the second term of Eq.~(\ref{PtLRB}) starts balancing out the first term, and gives rise to a \textit{steady growing} region.   A rough estimation of perfect state transfer time can be given by setting \(\epsilon-\frac{1}{4}\epsilon^2=1\Rightarrow \epsilon=2\), which gives
\begin{equation}
t \approx L/v.
\label{LRP}
\end{equation}
The main weakness of  the Lieb-Robinson method is the lack of dependence on the specific form of Hamiltonian and the circuit depth.  Nevertheless, it  offers useful insights into the difficulty of state transfer problems. In the later numerical section, we confirm the existences of the exponentially suppressed region, the exponentially growing region~(see \cref{p46-20}), the steady growing region, and the linear dependence between $t_f$ required for state transfer and the number of qubits $N$~(see  \cref{tf-N-2-19}). 

%However, there may be a constant gap between the Lieb-Robinson velocity and the actual quantum speed limit in our system because we are only considering single excitation subspace. In such single excitation subspace, the maximum group velocity is $v=2J$ (can be solved from the spectrum), which may be more relevant for our QAOA numerical result.

\section{Numerical Optimization of the QAOA}\label{NumericalSection}

Our analytic  success probability versus circuit depth scaling analyses    so far do not assume the optimality of the QAOA solution. To verify the tightness of these results for optimized QAOAs, we explore in this section the numerically optimized QAOA performance in regard to its success probability scaling as a function of the circuit depth and the physical runtime. We start by introducing briefly our numerical optimization method, and then describe and analyze the optimized QAOA performance obtained from our numerical method. We show that the quadratic Grover-like speedup shown in our analytic spectral analysis is also present in the numerically optimized QAOA solutions. We also show that when the circuit depth is too low, the given QAOA protocol becomes uncontrollable: its control landscape no longer possesses only global optimal but also   many local optimum points. Consequently, the optimized QAOA no longer necessarily guarantees the existence of a high fidelity state transfer scheme.  This finding unveils the relation between the $F-p$ scaling and the controllability of the underlying physical system.

We optimize QAOA parameters under two different constraints, one with limited physical run time and the other one with unlimited physical run time. Both situations are experimentally relevant. If the coherence time is sufficiently large, we may want to achieve state transfer with a minimum number of switches between $\hat{H}_B$ and $\hat{H}_C$. In contrast, if the switching operation is easy, the total physical run time should be minimized.  The practical implementation of QAOAs on near-term quantum computing hardware  is outside the scope of this article.

%\subsection{Analysis on depth-1 case}

\subsection{Numerical methods}\label{numerical-methods}

% introducing why this numerical work is hard, and what methods have been tried and failed

We use a gradient descent method to numerically determine the maximum achievable fidelity given $t_f$ and $p$.
We choose the optimization parameters as the duration for each QAOA Hamiltonian evolution $\delta_B^k$ and $\delta_C^k$ of each $k$th iteration with $k\in [p]$ for a depth-$p$ QAOA. The total physical runtime is the sum of all QAOA iterations: $t_f=\sum_k(\delta_B^k+\delta_C^k)$.  
We chose the parameter ranges for the physical run time $t_f$ and the circuit depth $p$   by preliminary numerical experiments. \Cref{table-tfp} and \cref{table-tfp2} summarize the parameters we performed grid search on. To increase the reliability of  the total  number of random restarts for gradient descent iterations, we ran preliminary experiments with a varying size of random uniform (over 1-simplex) initial conditions, and discovered that 200 random initial conditions were enough for $N=2\to15$ qubits cases to find approximate global optimal solutions. For $N=16\to20$ qubits, we increased the number of random restarts to 400. We use  L-BFGS Matlab toolbox for the QAOA optimization, which can be efficiently parallelized for large-scale experiments. These numerical results to be discussed below are summarized in \Cref{summary-n-r}.

\subsection{Numerical results for unlimited $t_f$ }\label{No-limit-tf}

We start with the optimized QAOA performance when the physical run time $t_f$ is not fixed. The analytic result at low circuit depth limit (Eq.~\eqref{Pp-low-depth}) coincides with our numerically result in \cref{subsecpp}. And the connection between QAOA circuit depth and the controllability is demonstrated in the numerical results on the control landscape in \cref{subsecControlLands}.

\subsubsection{Maximum achievable fidelity versus circuit depth p}\label{subsecpp}
\begin{figure}[h]
\subfloat[$N=2\to10$ qubits]{\includegraphics[width=0.95\linewidth]{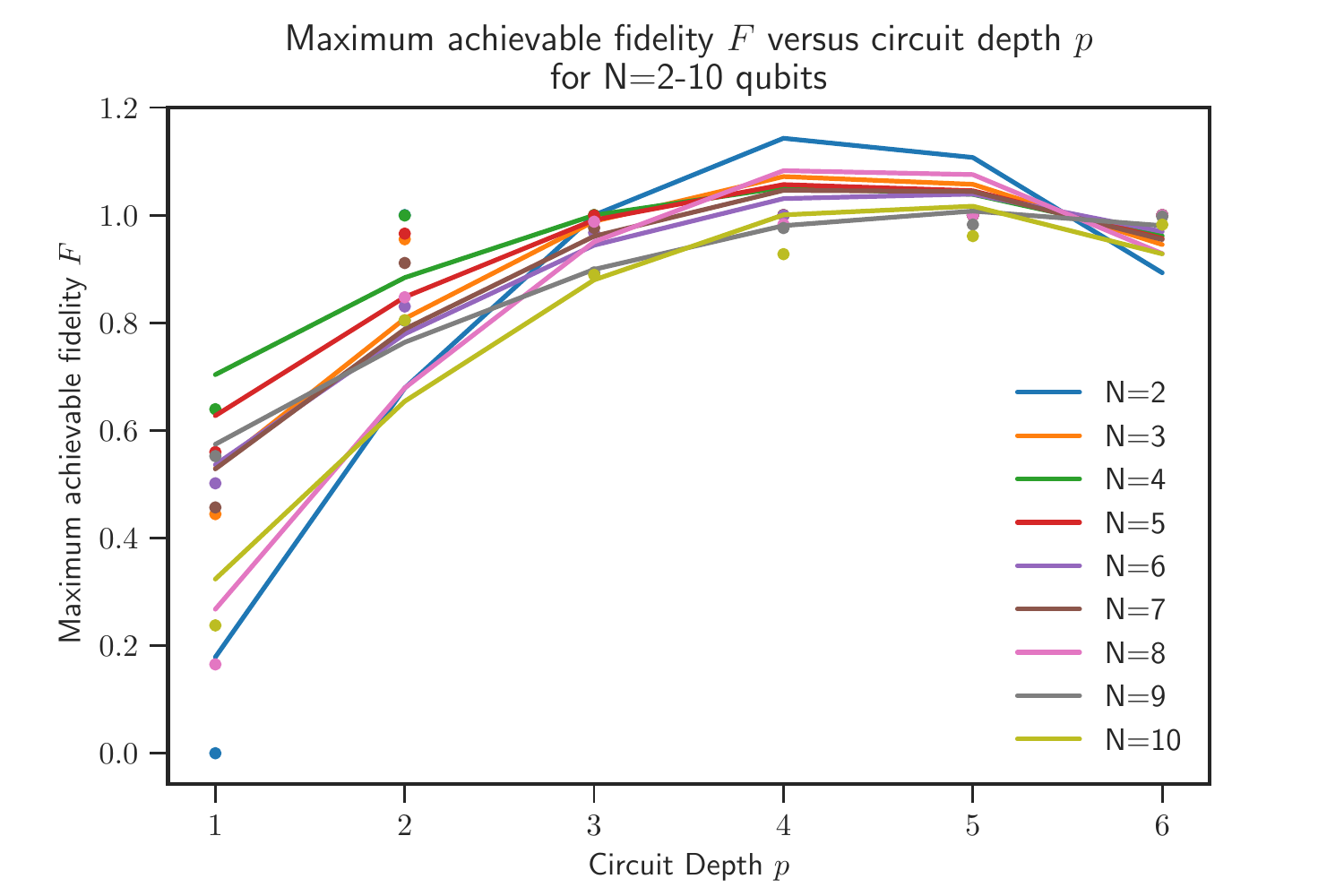}}

\subfloat[$N=11\to20$ qubits]{\includegraphics[width=0.95\linewidth]{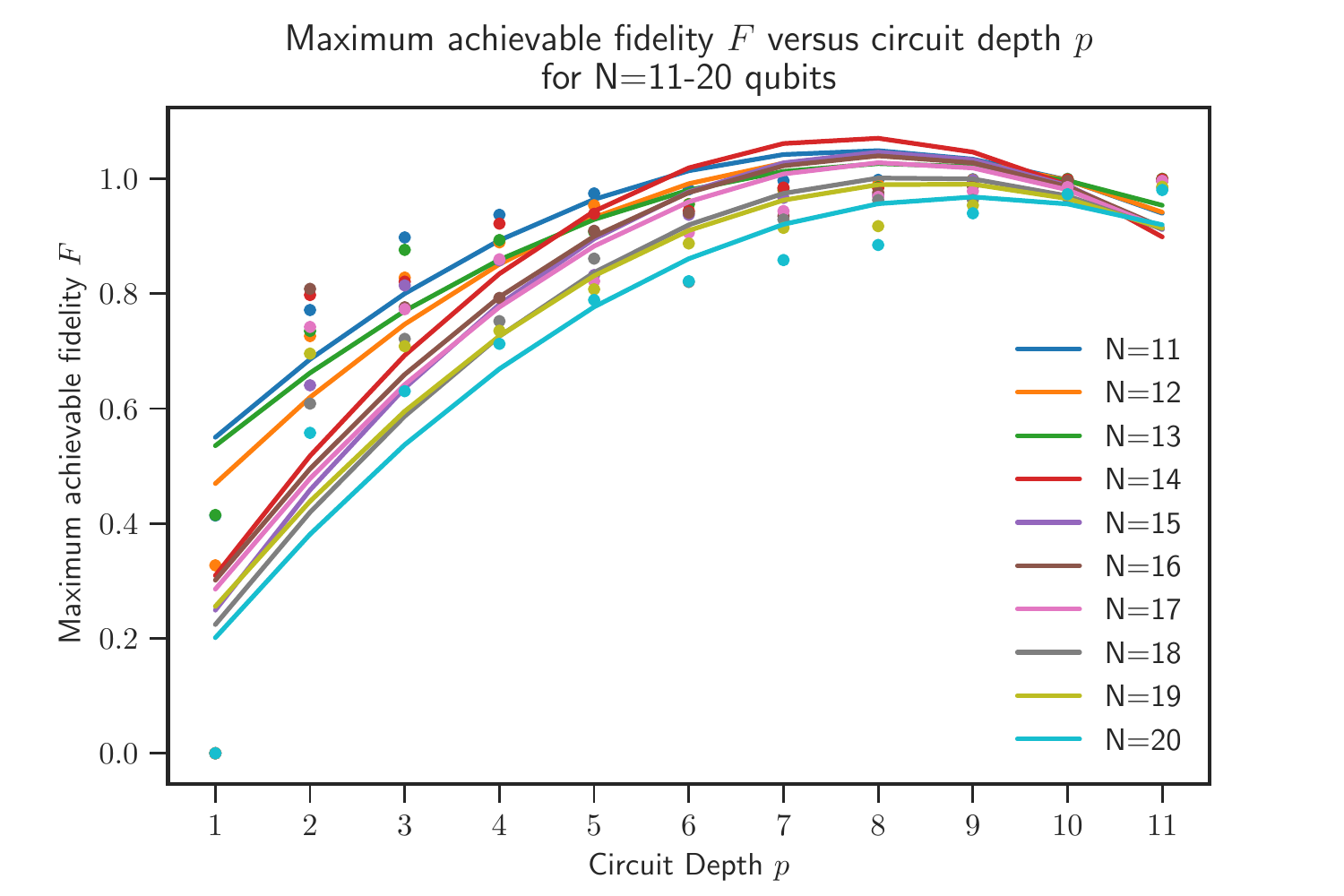}}
\caption{Maximum achievable fidelity $F$ using QAOAs as a function of the circuit depth $p$ with no constraints on $t_f$ for $N=2\to20$ qubits. The dots are numerical points. We fit the results with quadratic function $F(p)=ap^2+bp+c$, as represented by the lines. We  observe that the fidelity grows at low circuit depth, and then slowly converges to 1.0.}
\label{fit2-20-quadratic}
\end{figure}

\begin{figure}[h]
\includegraphics[width=0.8\linewidth]{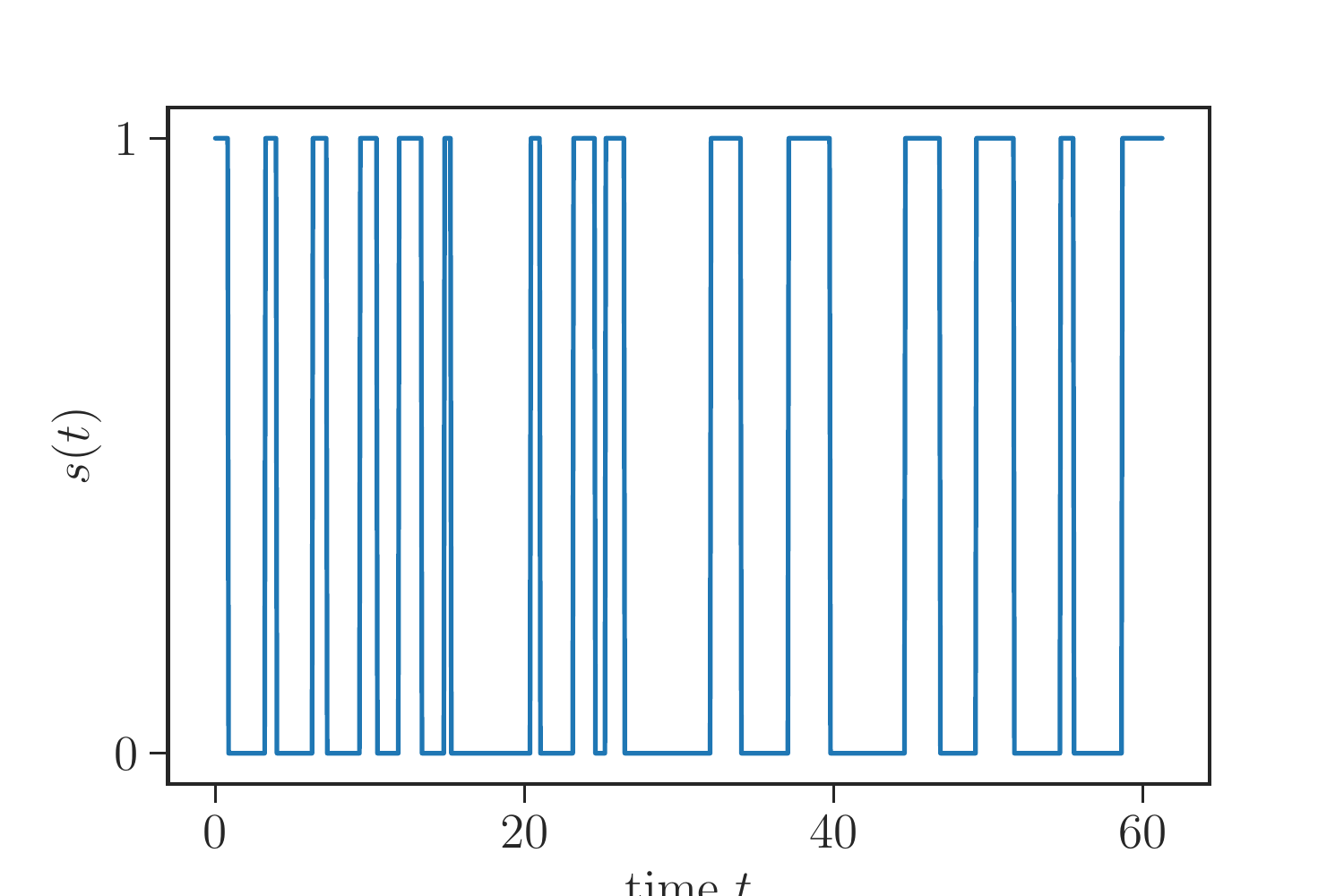}
\caption{An optimal bang-bang solution obtained through numerical optimizations for $N=10$ qubits and circuit depth $p=15$. In this case,  the optimal solution favors much shorter duration for the evolution under $\hat{H}_B$ than that under $\hat{H}_C$. This is in accord with our analytical result in Sec.~\ref{SuccessScalingSection}.} 
%favors much shorter duration for the evolution under $\hat{H}_B$ and almost uniform evolution under $\hat{H}_C$. 
\label{N10}
\end{figure}

\Cref{fit2-20-quadratic} shows maximum achievable fidelity $F$ as a function of circuit depth $p$ with no constraints on $t_f$ for $N=2\to 20$ qubits. 
The circuit depth dependence of success probability in \cref{fit2-20-quadratic} agrees with our analytic result: at the very beginning the quadratic dependence dominates (Eq.~(\ref{Pp-low-depth})), and a while later the exponential slow down dominates (Eq.~(\ref{PtLRB})).    
The time duration ansatz used  in our analytical result in Sec.~\ref{SuccessScalingSection} is also supported in our numerically optimized solution; see ~\cref{N10} for example,  where the intervals for $s(t)=1$   corresponding to the evolution under $\hat{H}_C$ is shorter on average than the duration in which $s(t)=0$.

\begin{figure}[h]
\subfloat[$N=2\to10$ qubits]{\includegraphics[width=0.95\linewidth]{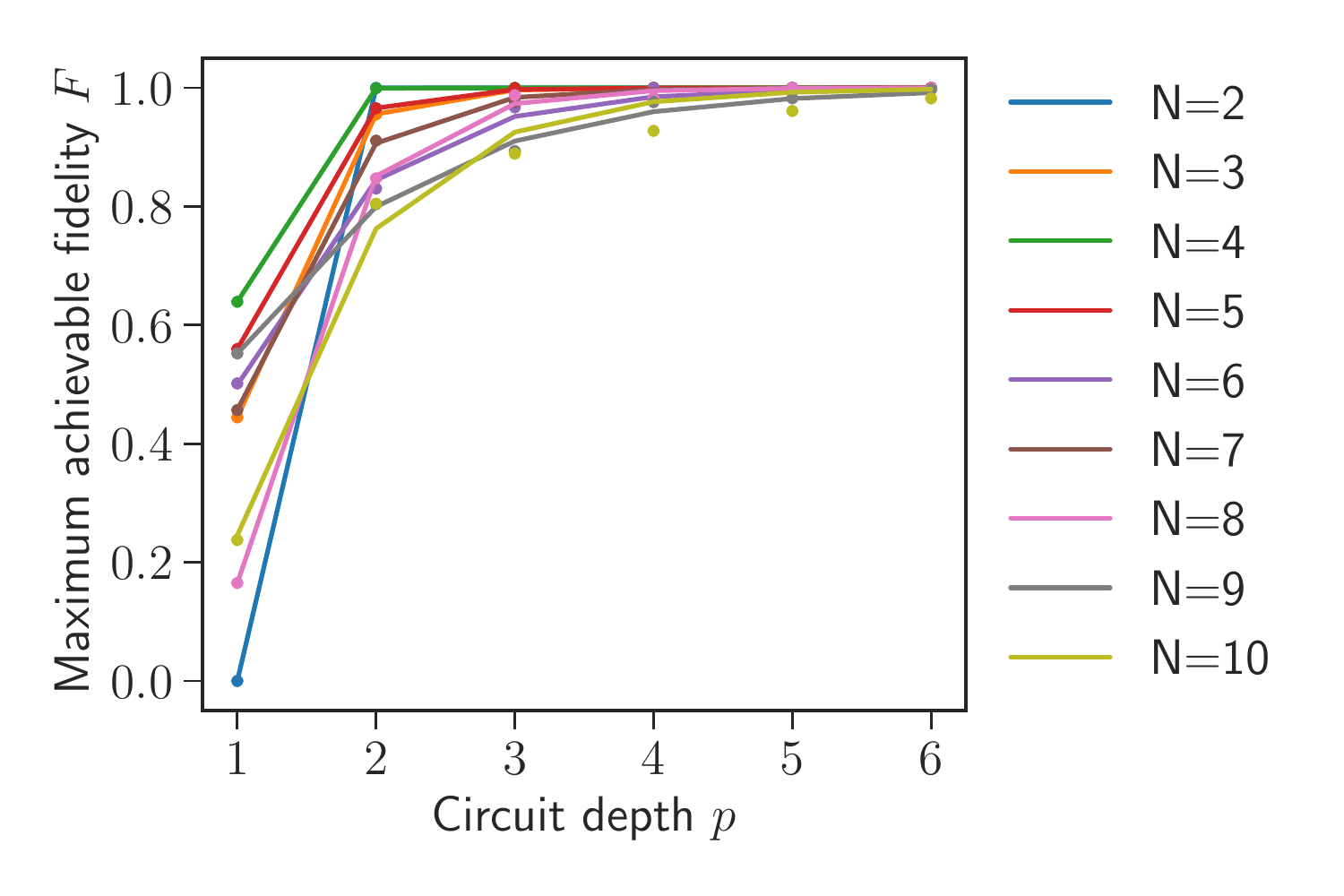}}

\subfloat[$N=11\to20$ qubits]{\includegraphics[width=0.95\linewidth]{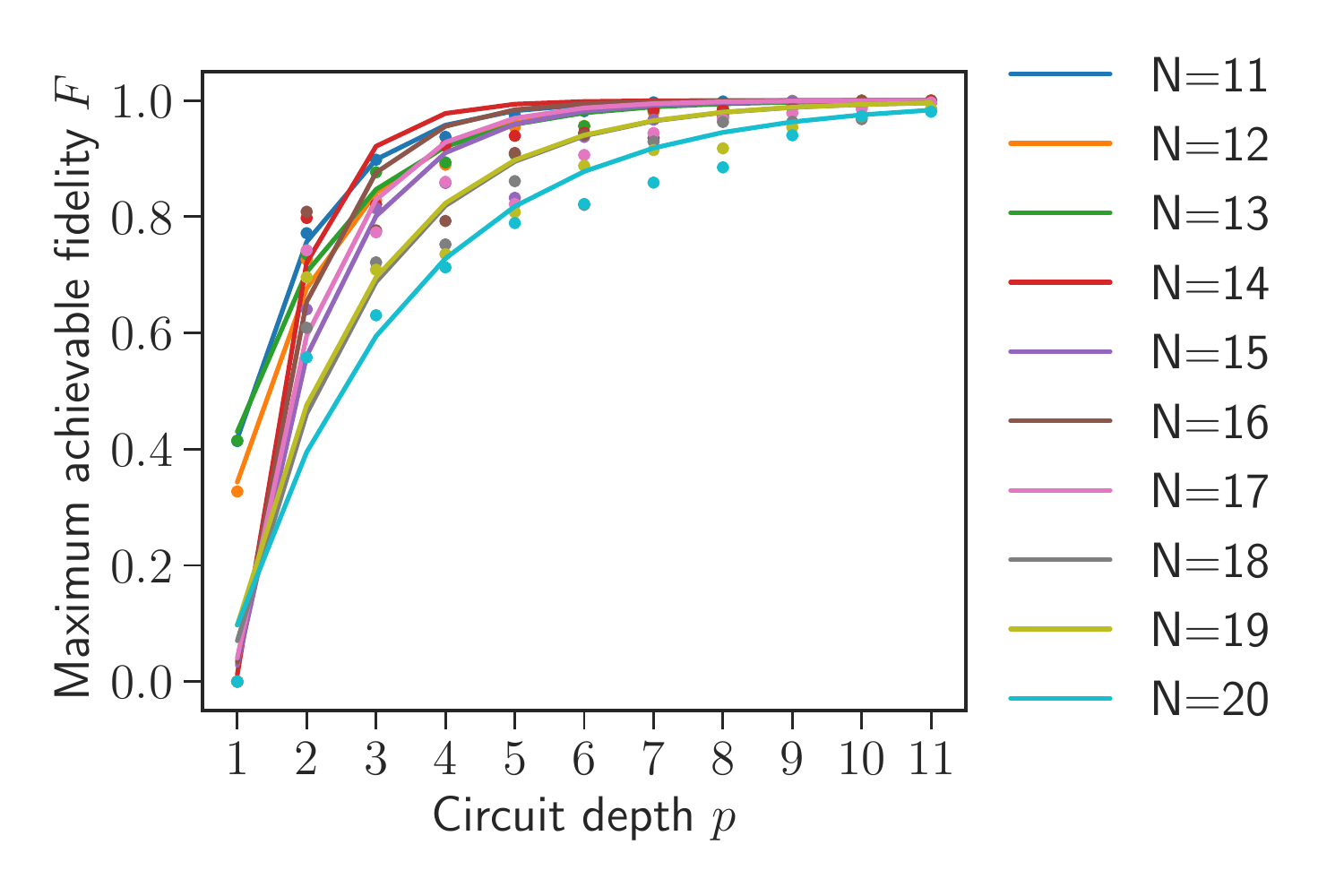}}
\caption{Maximum achievable fidelity $F$ as a function of the circuit depth $p$ with no constraints on $t_f$ for $N=2 \to 20$ qubits.The dots are numerical points.  We fit the results with inverted exponential function $F(p)=1 - \exp(-a(p - b))$. We can observe fidelity grows rapidly at low circuit depth, and then the fidelity slowly converge to unity.} 
\label{fit2-20}
\end{figure}

% \begin{table}[ht]
% \caption{Coefficient of determination $r^2$ of the fitting \cref{fit2-20}.  The fitting is good for 2-10 ($>$0.99). But not that good for 11-20. }

% \label{r2}
% \begin{tabular}{|l|l|l|l|l|l|l|l|l|l|l|}
% \hline
% $N$     & 2     & 3     & 4     & 5     & 6     & 7     & 8     & 9     & 10    & 11    \\ \hline
% $r^2$ & 1.000 & 0.999 & 0.999 & 0.999 & 0.996 & 0.998 & 0.999 & 0.995 & 0.983 & 0.997 \\ \hline
% $N$     & 12    & 13    & 14    & 15    & 16    & 17    & 18    & 19    & 20    &       \\ \hline
% $r^2$ & 0.988 & 0.980 & 0.970 & 0.965 & 0.910 & 0.921 & 0.938 & 0.891 & 0.938 &       \\ \hline
% \end{tabular}
% \end{table}

%

%
%\subfloat[$p=3$]{\includegraphics[width=0.3\linewidth]{figs/nl/example-solution/N10/N10P3.pdf}}
%\subfloat[$p=9$]{\includegraphics[width=0.3\linewidth]{figs/nl/example-solution/N10/N10P9.pdf}}

%
%\begin{figure}[h]
%\subfloat[$p=7$]{\includegraphics[width=0.3\linewidth]{figs/nl/example-solution/N20/N20P7.pdf}}
%\subfloat[$p=17$]{\includegraphics[width=0.3\linewidth]{figs/nl/example-solution/N20/N20P19.pdf}}
%\subfloat[$p=27$]{\includegraphics[width=0.3\linewidth]{figs/nl/example-solution/N20/N20P27.pdf}}
%\caption{Example solutions for $N=20$} 
%\label{N20p}
%\end{figure}
%

% write a transition paragraph here?

\subsubsection{Control landscape}\label{subsecControlLands}
\begin{figure}[h]
\subfloat[]{\includegraphics[width=0.95\linewidth]{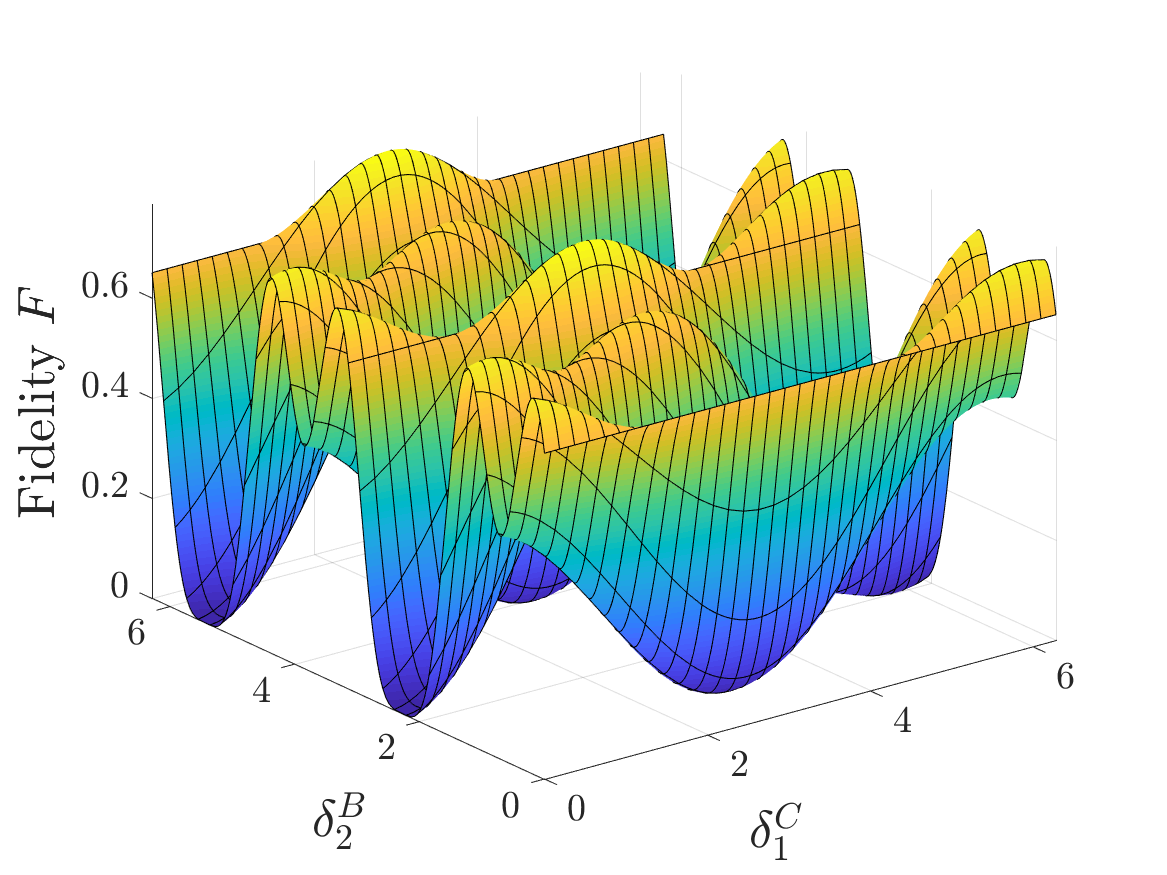}}

\subfloat[]{\includegraphics[width=0.95\linewidth]{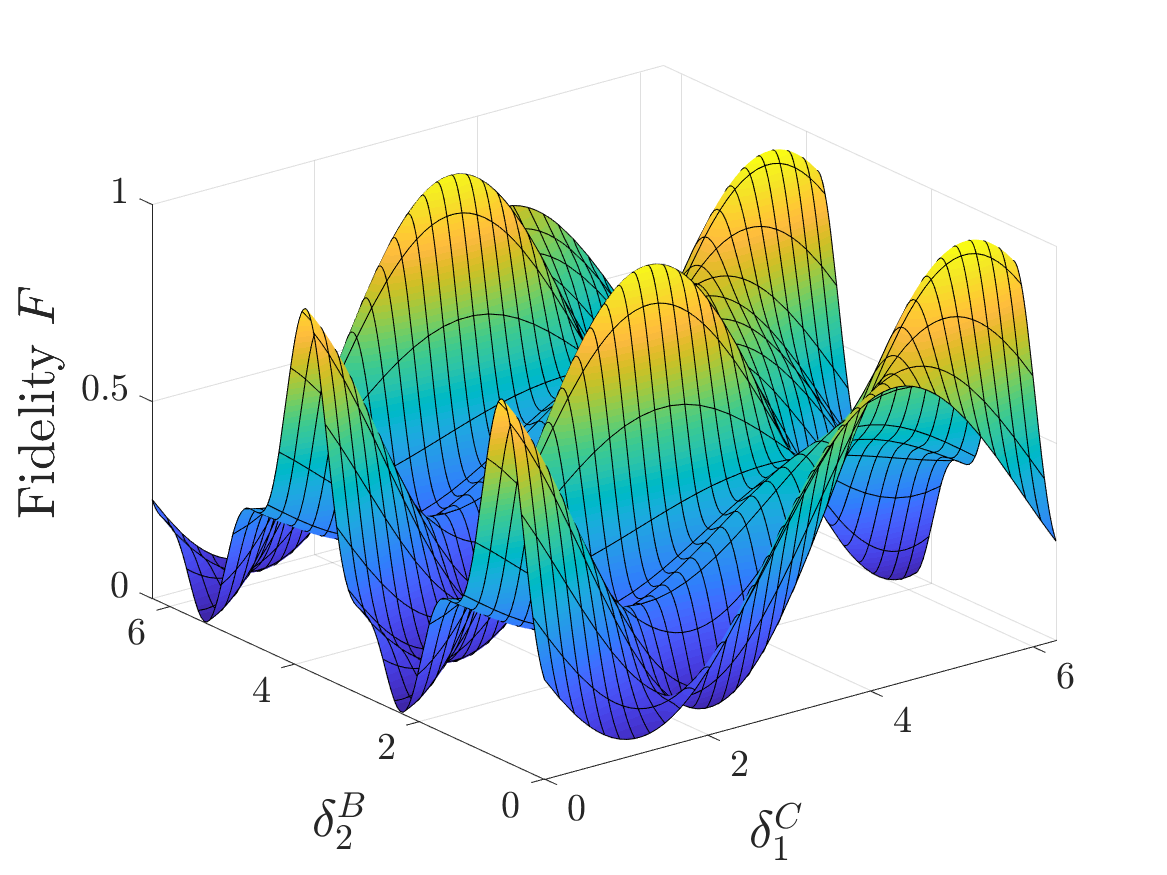}}
\caption{A comparison of the control landscapes of the QAOA with low circuit depth $(p=2)$ and of that with a larger circuit depth $(p=4)$ for a three-qubit system. (a) The control landscape for two chosen variables of a depth-2 QAOA, which admits a maximum achievable fidelity of 0.787. As we observed many local minima, the system is uncontrollable. (b) The control landscape for two chosen variables of a depth-4 QAOA, which admits a  maximum achievable fidelity of 1.000. Since all local minima are global minima, the system is controllable. }
\label{landscape}
\end{figure}

The optimization of QAOA can be regarded as a quantum control problem, where the durations of different QAOA Hamiltonian evolutions are the control parameters, and the fidelity of the state transfer is the control cost function to be maximized. Under this analogy,
when the system is controllable, the landscape of the control cost function over parameter space generically has only global minima \cite{rabitz2004quantum,PhysRevA.86.013405,russell2016quantum}. When the system is uncontrollable, the quantum control landscape admits many local minima \cite{PhysRevA.83.062306}. In our case, if we allow the circuit depth to be infinite, the system is controllable; but the controllability for an intermediate number of circuit depths demands further investigation. As a simple example, we plot the control landscape for $N=3$ qubits in \cref{landscape} with $p=2,4$. \cref{landscape} shows that the QAOA ansatz with $p=2$ is uncontrollable, as there are many local minima. In contrast, the case $p=4$ admits only global minima (at least for the plotted part); thus more controllable.  We review the control viewpoint of optimizing QAOA in \cref{optimal-control-pontryagins-principle}.

% transition paragraph?

\subsection{Numerical results for fixed $t_f$: optimized QAOA}\label{numerical-results}

We now discuss the optimized QAOA performance with a fixed physical run time $t_f$. 
In \cref{success-probability-p-versus-the-number-of-switches-p}, we investigate the  controllability dependence on $t_f$. Second, in \cref{success-probability-p-versus-total-run-time-t_f}, we discuss the Lieb-Robinson type scaling emerged in the optimized QAOA performance. 
%\subsection{$F$ versus $p$}
\subsubsection{Maximum achievable fidelity $F$ versus circuit depth $p$} \label{success-probability-p-versus-the-number-of-switches-p}

\begin{figure}[h]
\subfloat[Plots for a small $t_f=6$.]{\includegraphics[width=0.95\linewidth]{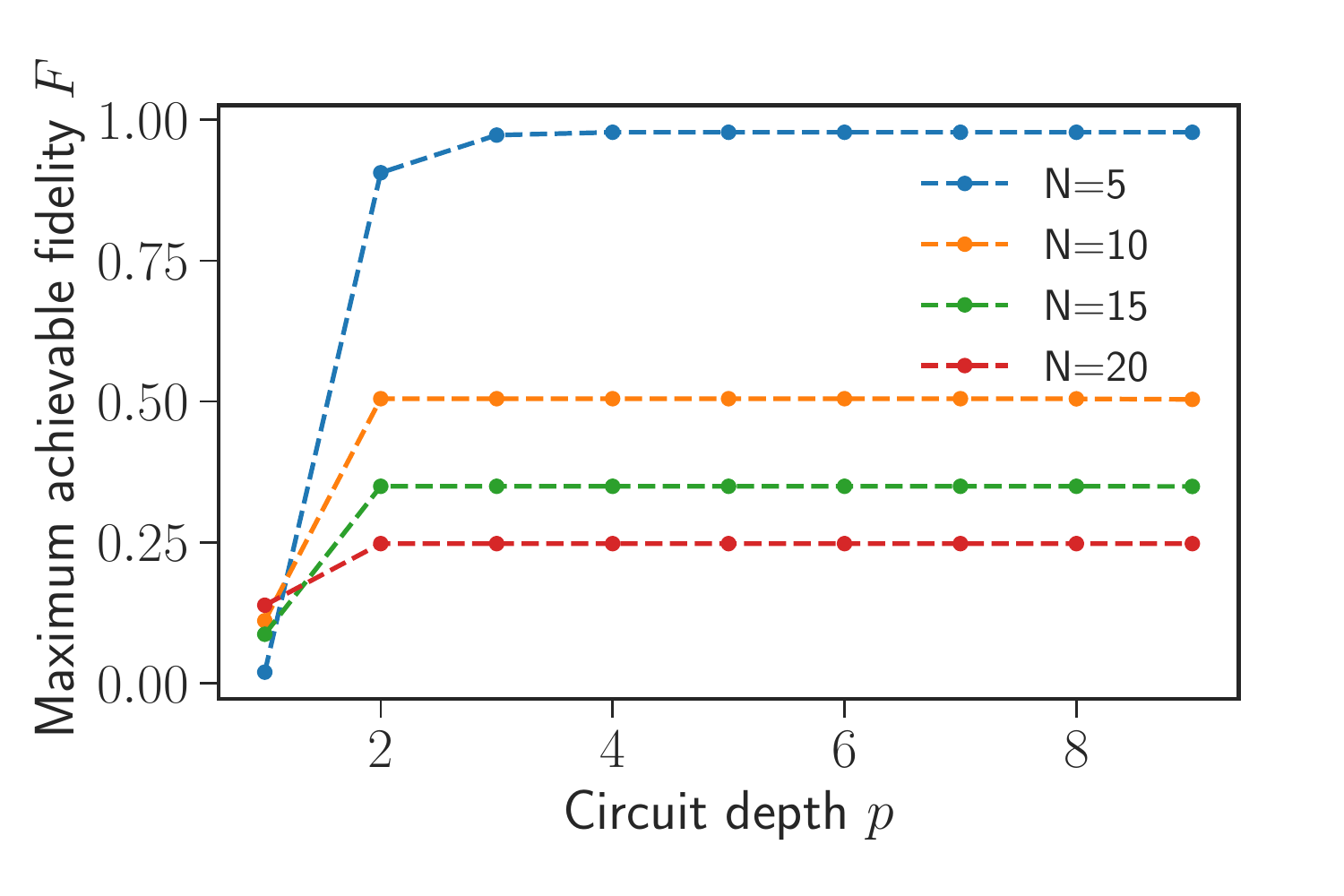}}

\subfloat[Plots for a medium $t_f=13$.]{\includegraphics[width=0.95\linewidth]{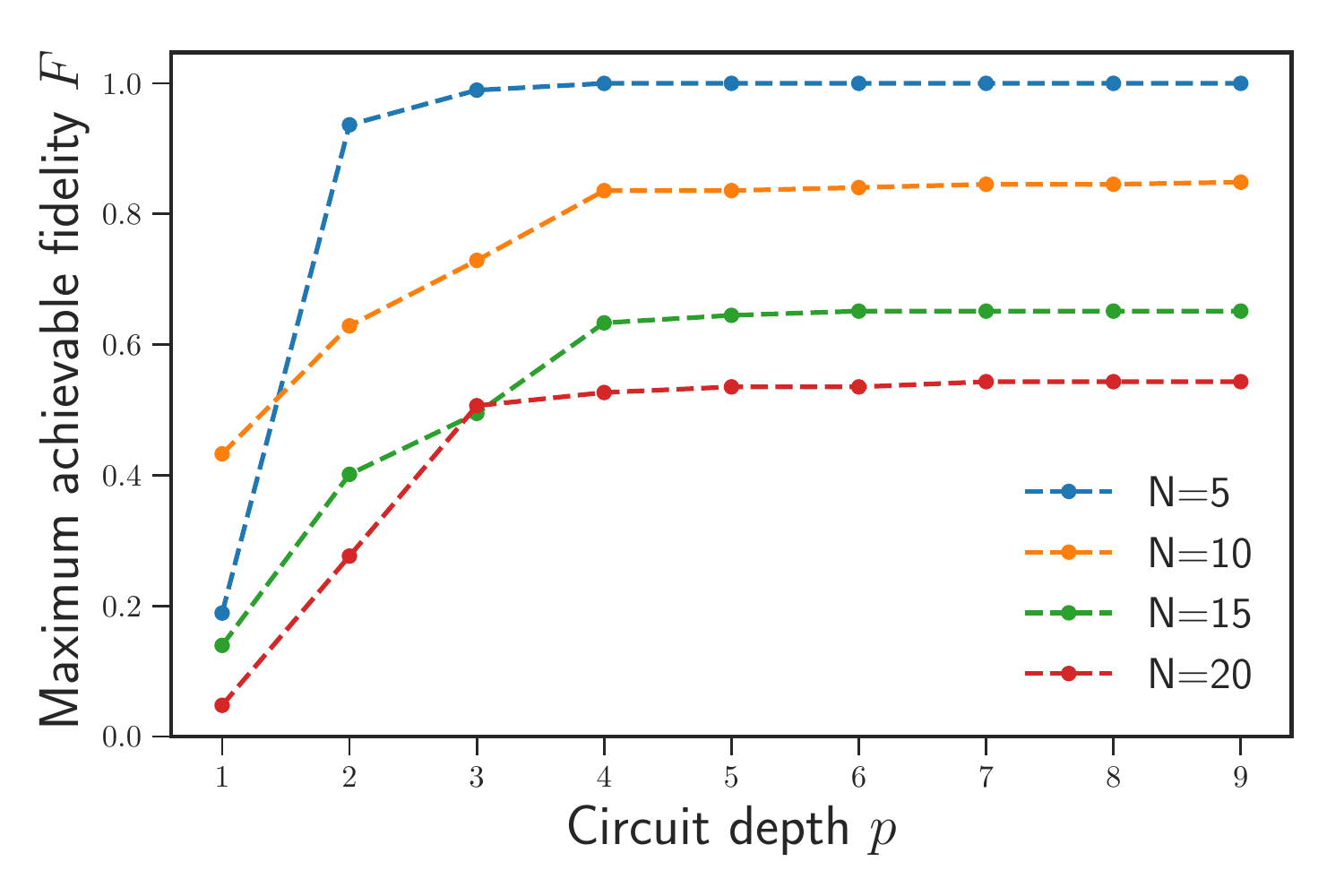}}
\caption{Maximum achievable fidelity $F$ versus circuit depth $p$ with the same fixed $t_f$ for $N=5,10,15,19$ qubits. For fixed $t_f$, we observed that there exists a circuit depth $p$ beyond which there will be no improvement of fidelity. We find a depth-3 circuit is sufficient for $t_f=6$ while a depth-4 circuit is needed for $t_f=13$.
}
\label{p6-13}
\end{figure}

\begin{figure}[h]
\subfloat[$N=2,4,6,8$ qubits. ]{\includegraphics[width=0.95\linewidth]{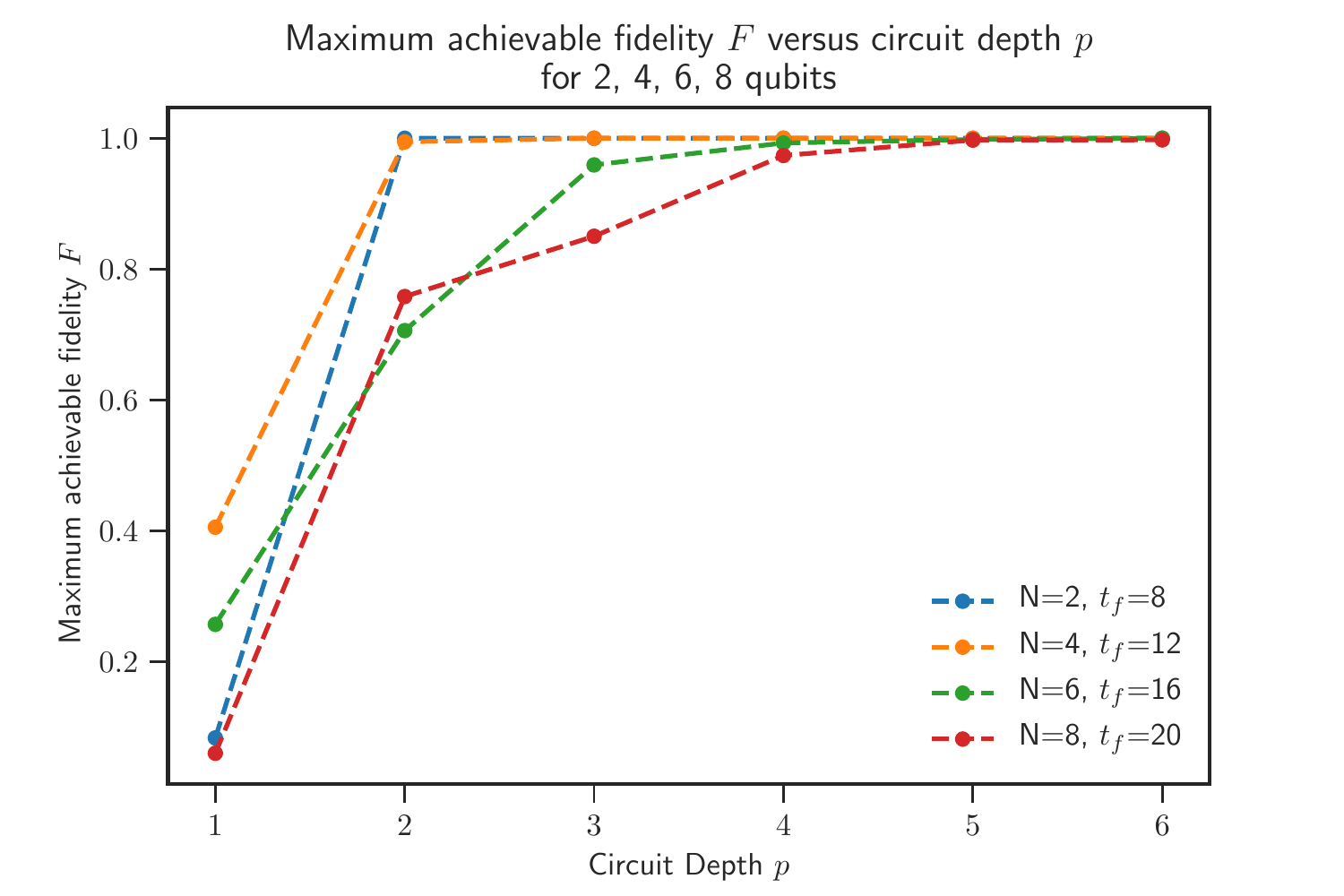}}

\subfloat[$N=10,12,14,16$ qubits.]{\includegraphics[width=0.95\linewidth]{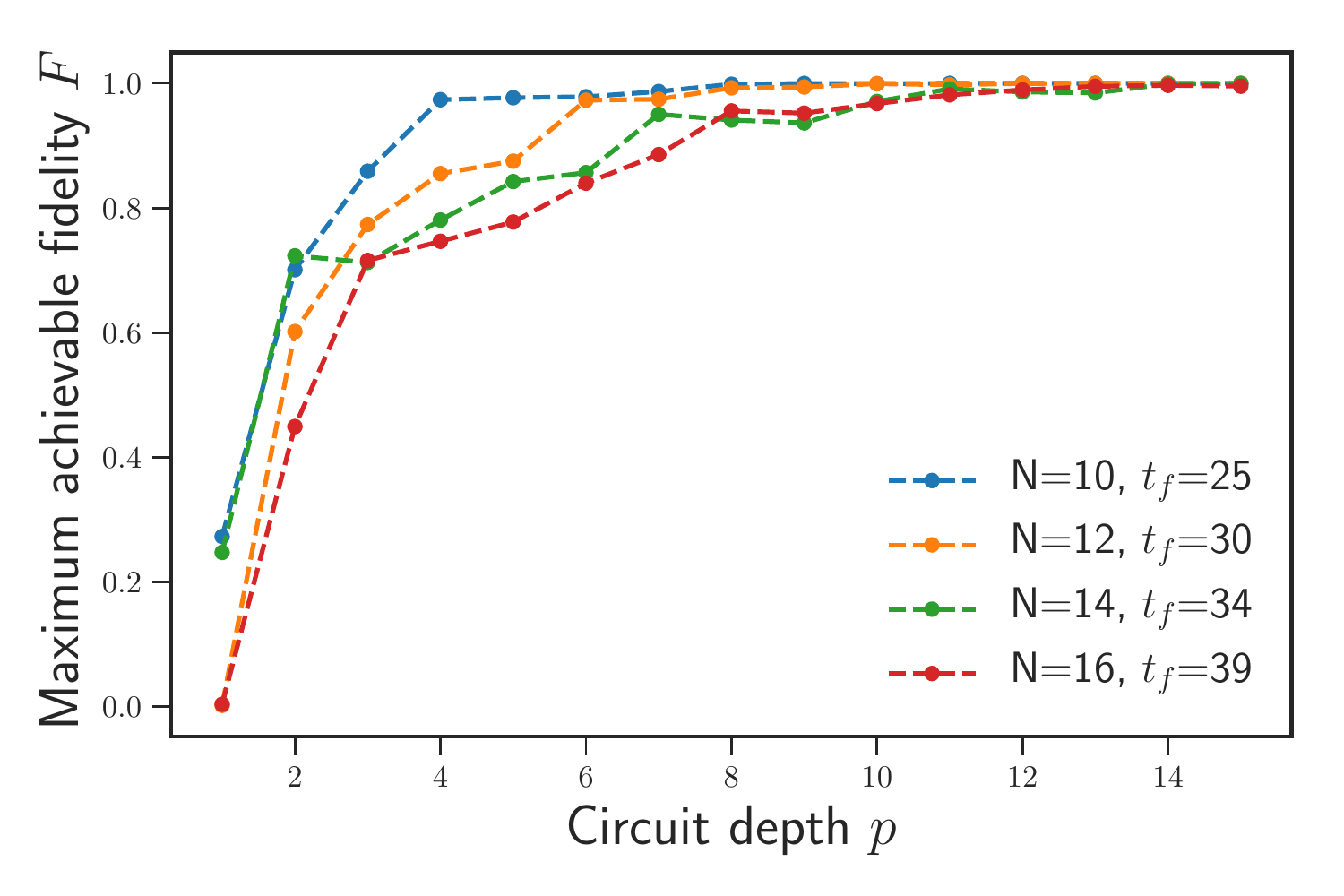}}
\caption{Maximum achievable fidelity $F$ versus circuit depth $p$ with fixed $t_f$. The physical run time $t_f$ is chosen through a hyper-parameter grid search separately.  }
\label{p2-16}
\end{figure}

In this section, we numerically study the maximum achievable fidelity $F$ versus the circuit depth $p$ for a  fixed $t_f$ in \cref{p6-13,p2-16,p2-16}.  Generally speaking, the larger circuit depth QAOA should always perform better than lower depth ones. However, if $p$ is too large, the difficulty of the QAOA optimization increases and the optimization can get stuck in local optima.  This results in a non-monotonic behavior in numerically optimized fidelity as a function of circuit depth.    For fixed $t_f$, there is a circuit depth $p$ beyond which fidelity can no longer be improved. As shown in \cref{p6-13}, for $t_f=6$ , the maximum achievable fidelity does not increase for circuit depth larger than $p=3$, and for $t_f=13$, the maximum achievable fidelity does not increase for circuit depth larger than $p=4$. This observation is also intimately related to the controllability of the QAOA: for the fixed run time, there exists a threshold circuit depth below which the QAOA is no longer controllable.

%In \cref{No-limit-tf}, we analyzed how many circuit depth $p$ is enough for perfect state transfer given unlimited time.   However, in reality, $t_f$ should be a practical, relevant quantity. As circuit depth, we observed there's a physical run time $t_f$ beyond which there will be no improvement of fidelity if the system is controllable. In \cref{p2-16}, we plot achievable fidelity $F$ versus circuit depth $p$ with fixed $t_f$ for different number of qubits. The similar slopes when $p$ is low may imply a suboptimal structure.

%
%Now in \cref{p10-15} we investigate the effect of increasing $t_f$ for a fixed number of qubits $N$. With a larger $t_f$, more circuit depth $p$ is needed for fidelity to converge.
%
%\begin{figure}[h]
%\subfloat[Achievable fidelity $F$ versus circuit depth $p$ with different $t_f$ for $N=10$ qubits. ]{\includegraphics[width=0.45\linewidth]{figs/tf3/plots/s10-1_4_7_10_13_16_19_22-200-8.pdf}}
%\subfloat[Achievable fidelity $F$ versus circuit depth $p$ with different  $t_f$ for $N=15$ qubits.]{\includegraphics[width=0.45\linewidth]{figs/tf3/plots/s15-1_5_10_15_20_25_30_35-200-12.pdf}}
%\caption{Achievable fidelity $F$ versus circuit depth $p$ with different $t_f$. With a larger $t_f$, more circuit depth $p$ is needed for fidelity to converge. The blue line($t_f=1$) is a signature for Lieb-Robinson suppression.}
%\label{p10-15}
%\end{figure}

%\subsection{$F$ versus $t_f$}
\subsubsection{Maximum achievable fidelity $F$ versus physical run time $t_f$}\label{success-probability-p-versus-total-run-time-t_f}

\begin{figure}[h]
\subfloat[Plots for $p=2$ (uncontrollable).]{\includegraphics[width=0.95\linewidth]{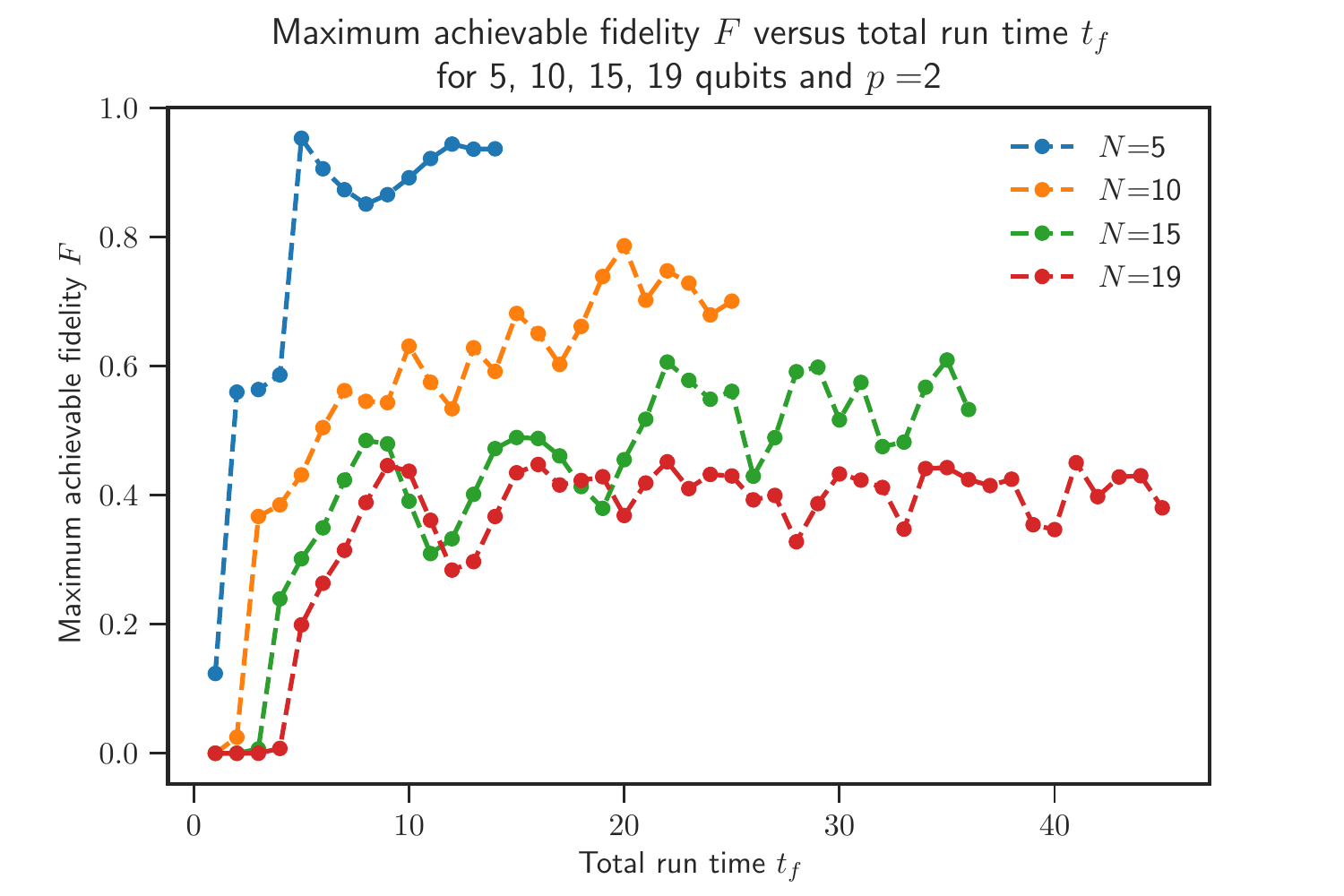}}

\subfloat[Plots for $p=9$ (controllable).]{\includegraphics[width=0.95\linewidth]{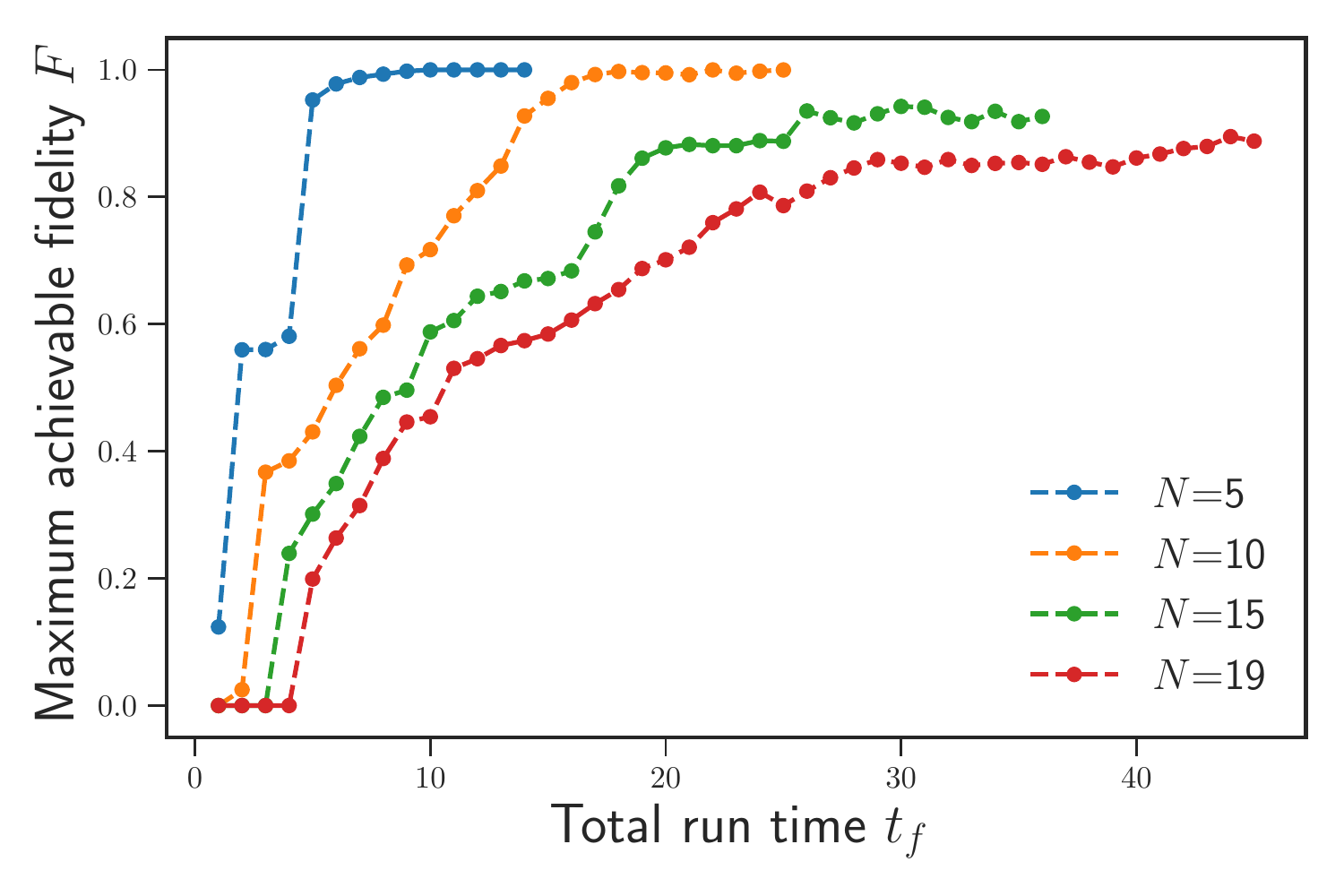}}
\caption{Maximum achievable fidelity $F$ versus physical run time $t_f$ with a fixed circuit depth $p$ for $N=5,10,15,19$ qubits. The oscillating behavior is due to the low circuit depth; such behaviour disappears when the circuit depth $p$ grows larger (equivalently, the system gets more controllable).}
\label{p2-9-}
\end{figure}

\begin{figure}[h]
%[Plots for $p=2$ (low).]
\subfloat{\includegraphics[width=0.95\linewidth]{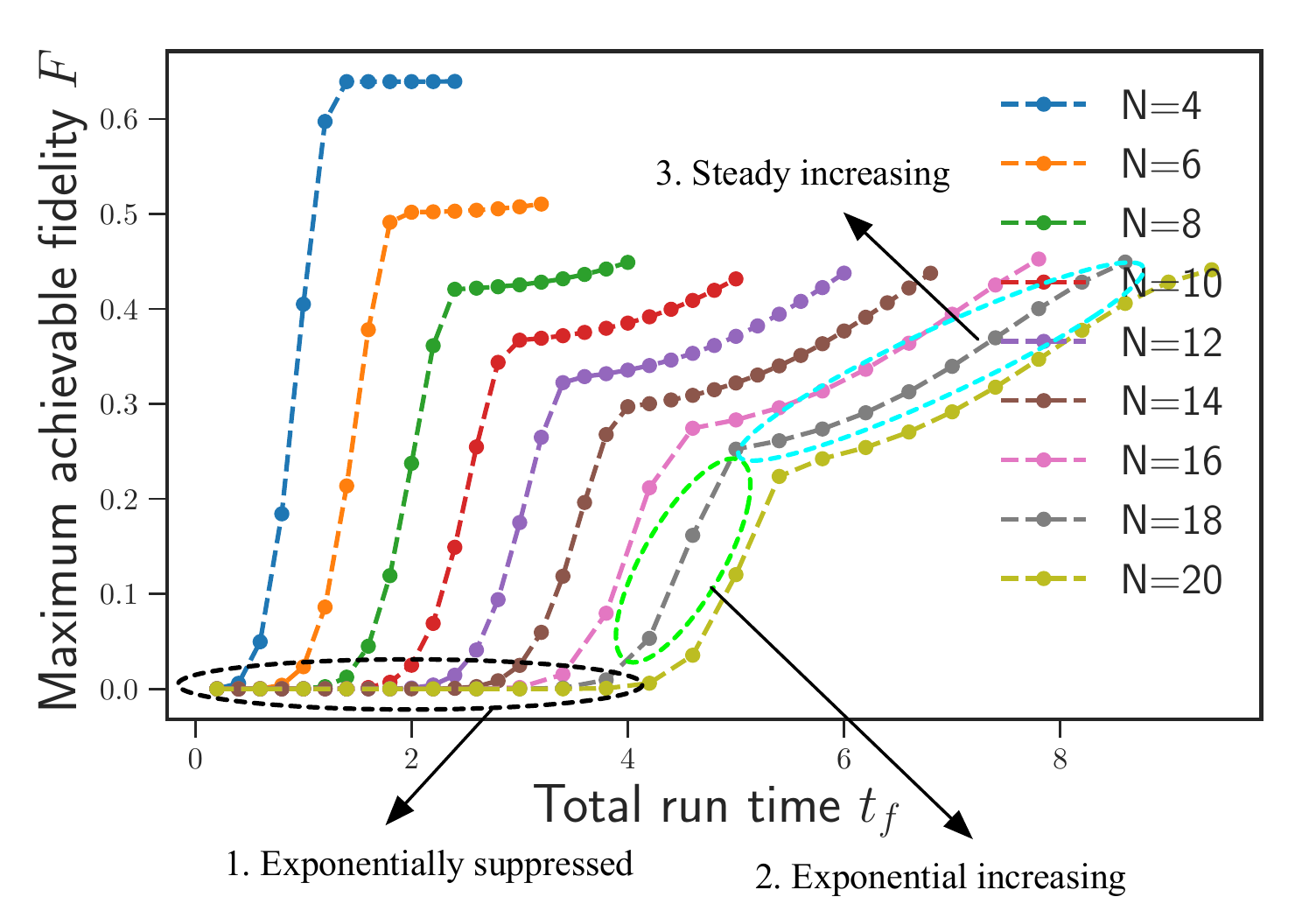}}
%\subfloat{\includegraphics[width=0.95\linewidth]{figs/tf/plots_smalltf/sptfmax/4_6_8_10_12_14_16_18_20-m1-200.pdf}}
\caption{Maximum achievable fidelity $F$ versus small physical run time $t_f$ with a sufficiently large circuit depth $p$ for $N=4,~6,~8,~10,~12,~14,~16,~18,~20$ qubits. In general, we identify three different growing patterns: (1) \emph{exponentially suppressed region;} (2) \emph{exponentially increasing region;} (3) \emph{steady increasing region.}}
\label{p46-20}
\end{figure}

\begin{figure}[h]
\subfloat[$N=2,4,6,8$ qubits.]{\includegraphics[width=0.95\linewidth]{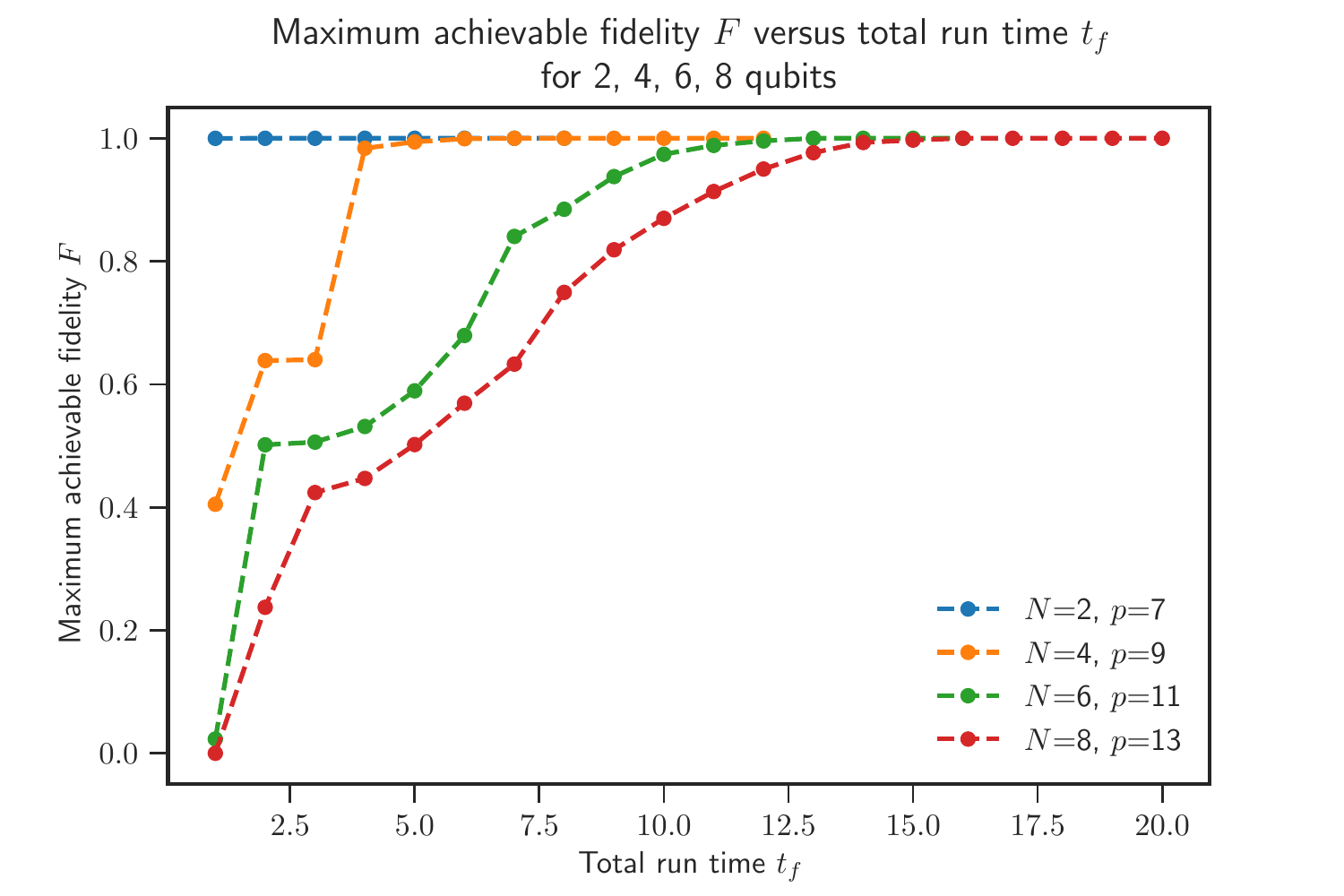}}

\subfloat[$N=10,12,14,16$ qubits.]{\includegraphics[width=0.95\linewidth]{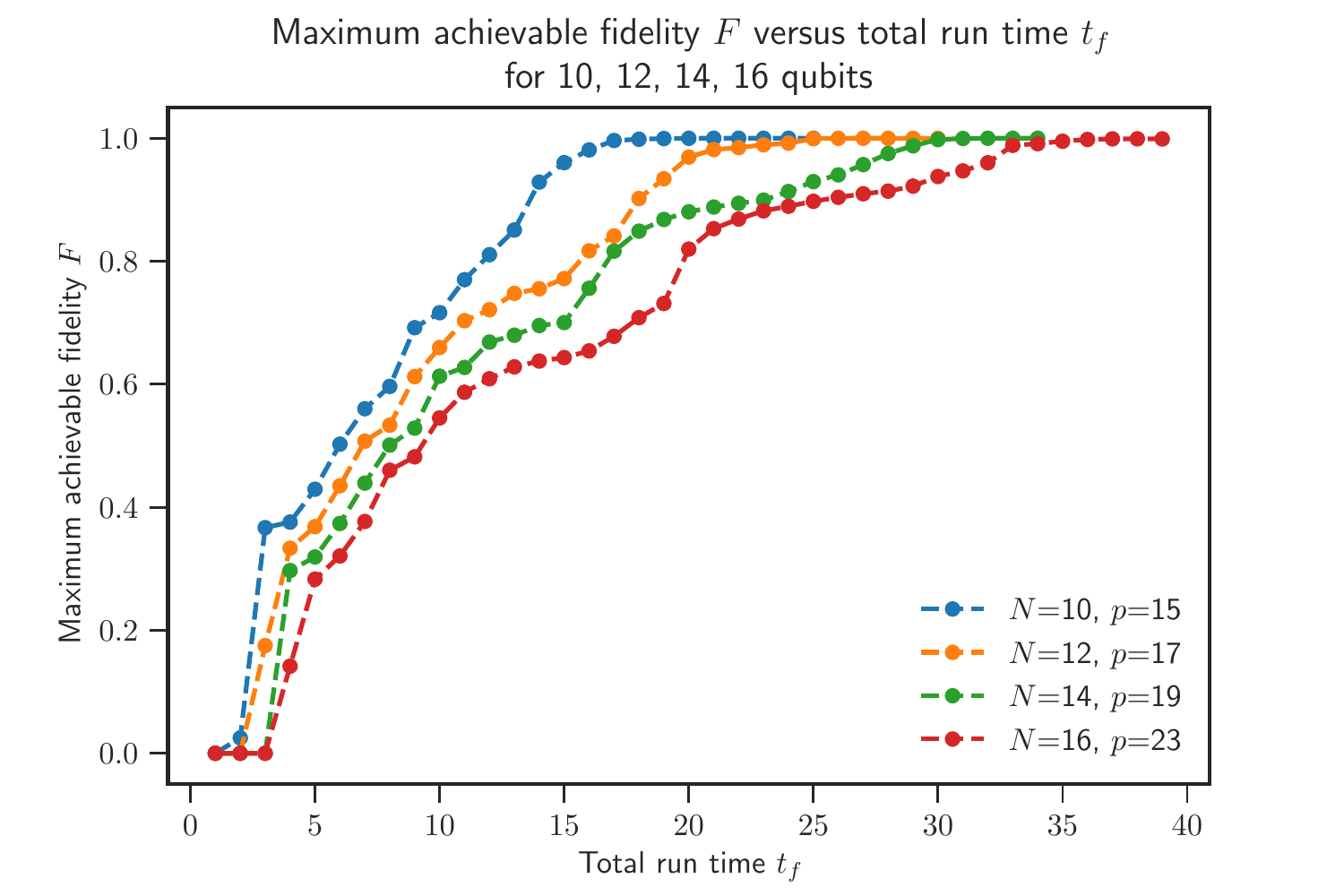}}
\caption{Maximum achievable fidelity $F$ versus physical run time $t_f$ with a sufficiently large circuit depth $p$.}
\label{2-16-m1}
\end{figure}

In this subsection, we investigate the performances of the QAOA with a fixed physical runtime in \Cref{p2-9-}. We identify three different temporal dependencies of fidelity as predicted by the Lieb-Robinson bound, as depicted in \cref{p46-20}: \emph{exponentially suppressed region}; \emph{exponentially growing region}; and \emph{steady growing region}.
We find that the longer the physical run time $t_f$ is, the better achievable fidelity will be under the condition that the circuit depth $p$ is sufficiently large and $t_f$ is outside of the highly suppressed region.  For a low depth circuit, the oscillating in success probability occurs (\cref{p2-9-}(a)),  which is a sign of uncontrollability. Such oscillation disappears for sufficiently large $p$,  see \cref{p2-9-}(b). 
The three regions of the growth region become more apparent as circuit depth increases, see in \cref{2-16-m1}. 

\begin{figure}[h]
\includegraphics[width=0.95\linewidth]{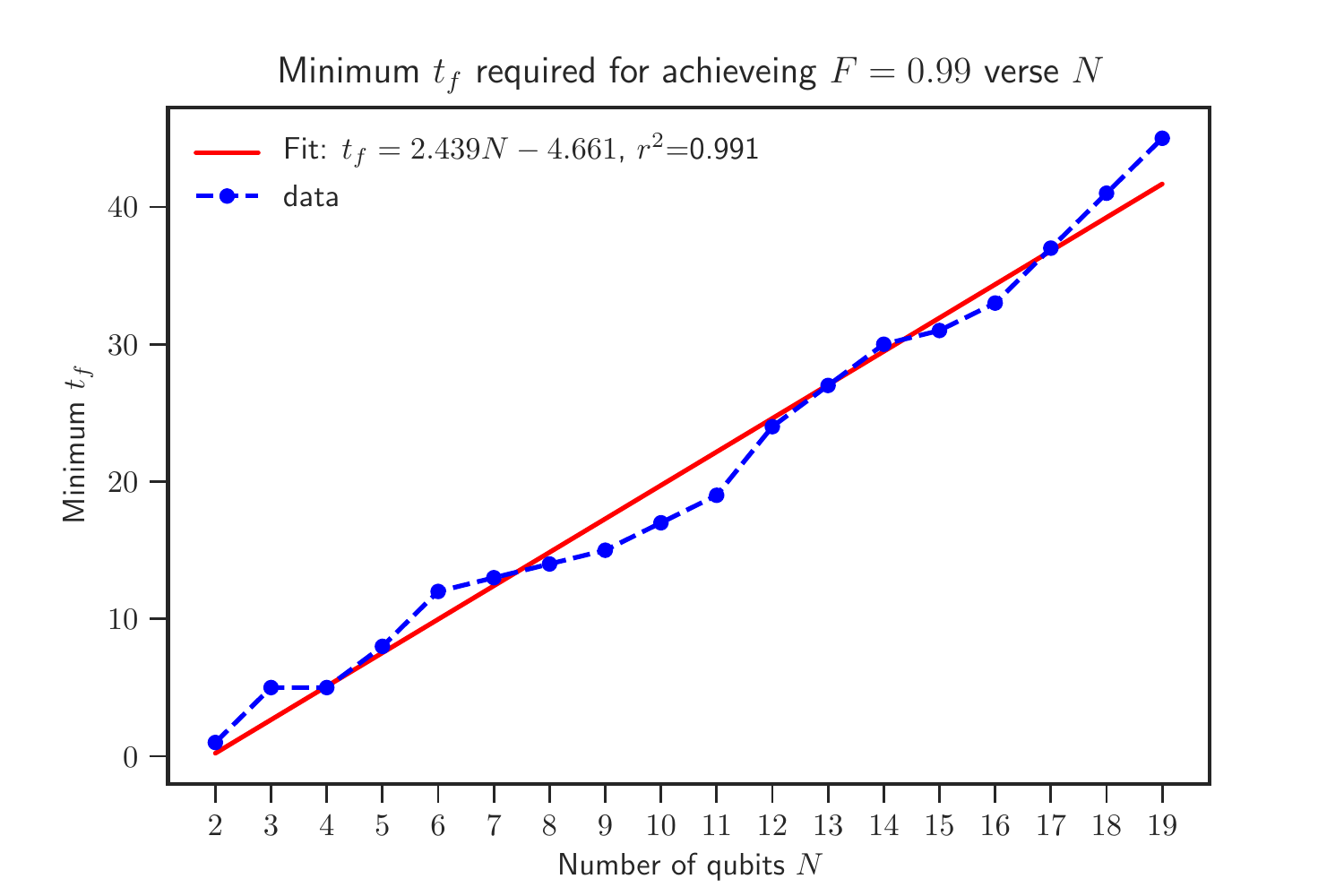}
\caption{Minimum required run time $t_f$ for achieving fidelity $F=0.99$ versus the number of qubits ($N=2\to 19$). The Lieb-Robinson bound gives a lower bound of approximately $t_f=0.03N+c$.}
\label{tf-N-2-19}
\end{figure}

\begin{figure}[h]
\subfloat[$N=10$ qubits. ]{\includegraphics[width=0.95\linewidth]{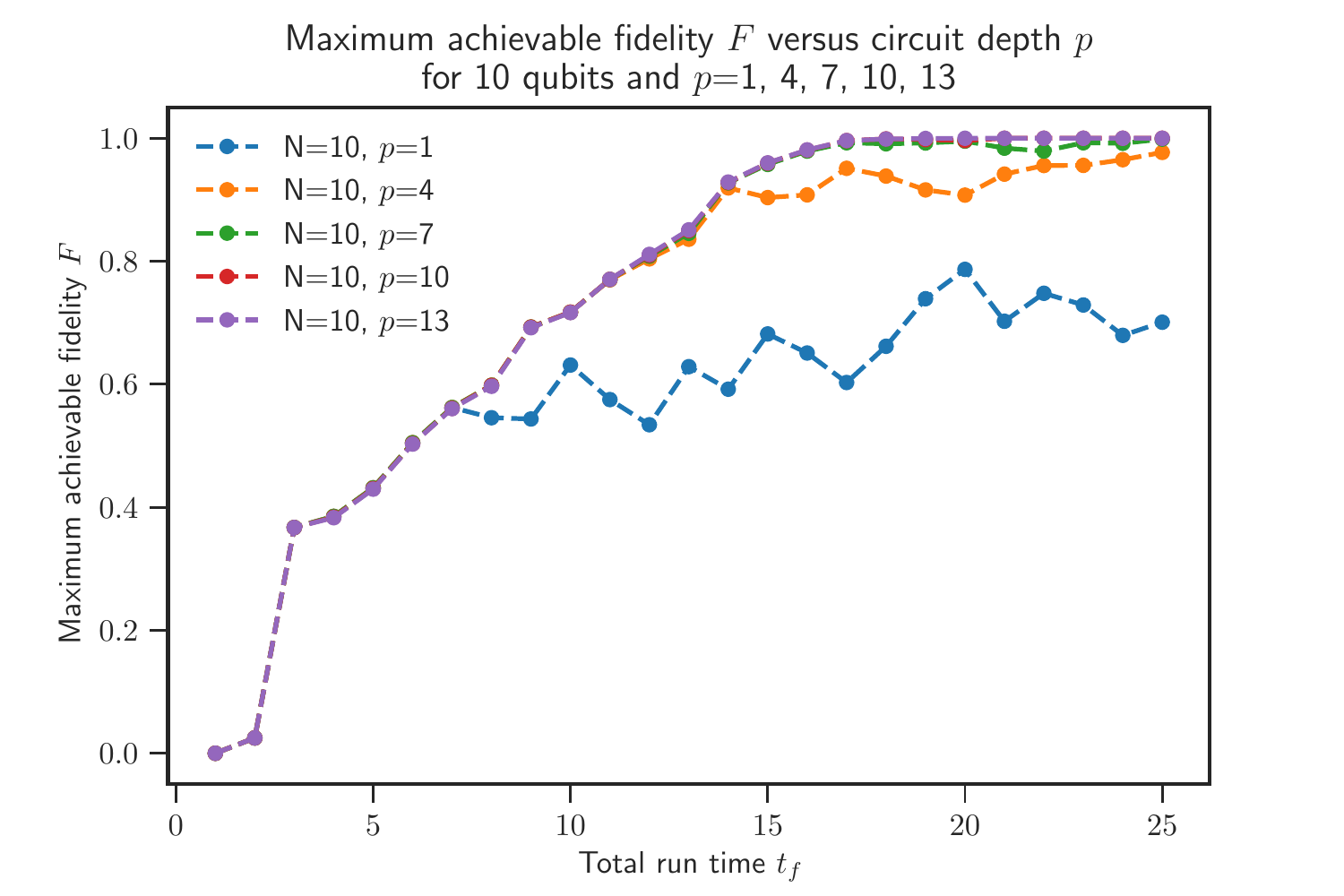}}

\subfloat[$N=15$ qubits.]{\includegraphics[width=0.95\linewidth]{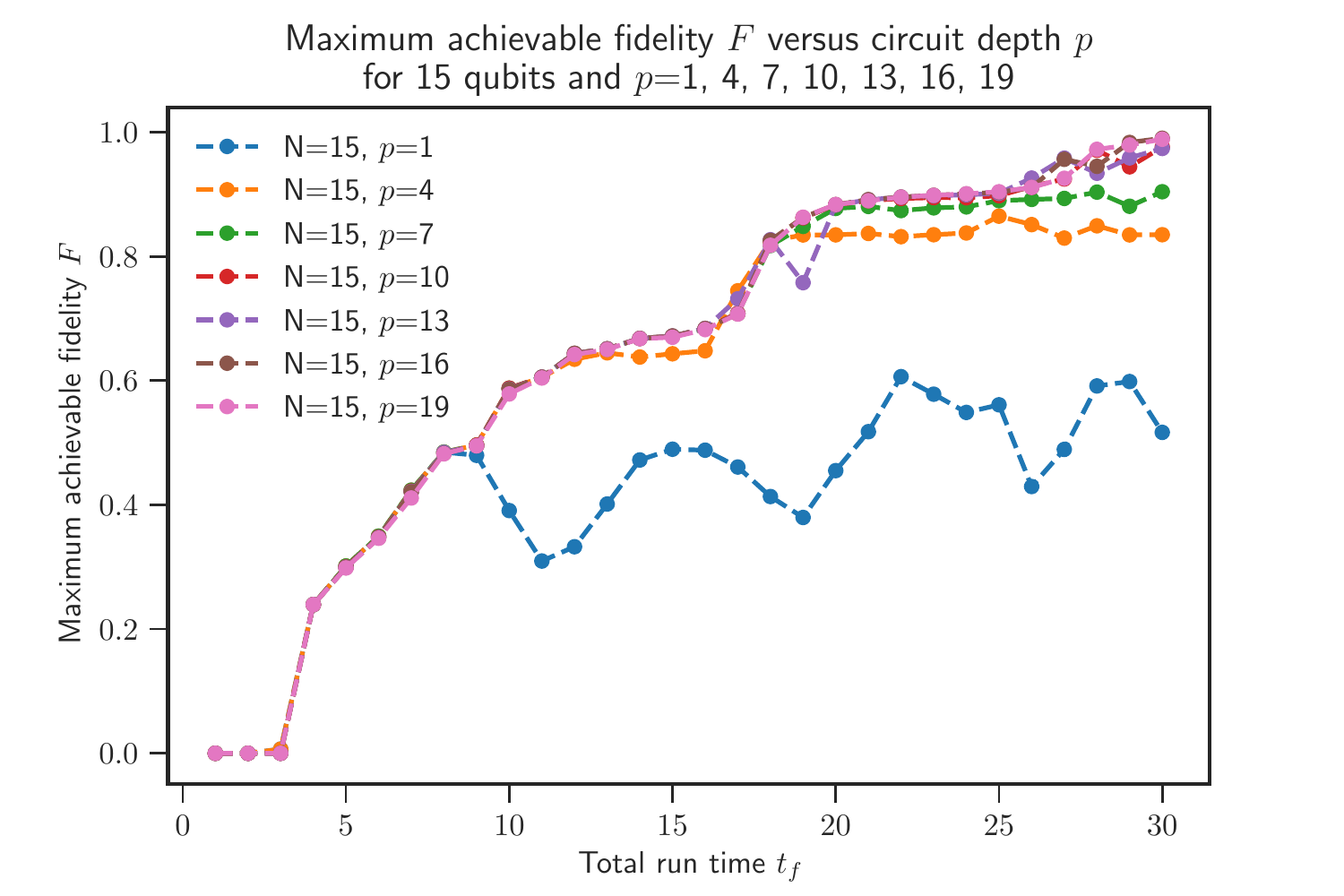}}
\caption{Maximum achievable fidelity $F$ versus physical run time $t_f$ using QAOA as a function of different different circuit depth $p$. The lines with oscillating behaviors are uncontrollable.}
% The oscillating behaviors signals the uncontrollable?
\label{10-15-F-tf}
\end{figure}

In \cref{tf-N-2-19}, we fit the minimum required run time $t_f$ for achieving fidelity $F=0.99$ as a function of the number of qubits ($N=2\to19$). The linear dependence from the Lieb-Robinson bound \cref{LRP} is seen with $t_f\sim 2.439N$. Given the same amount of run time, we show in \cref{10-15-F-tf}  that the  QAOA with a higher circuit depth necessarily achieves higher success probability. 

\begin{figure}[h]
\includegraphics[width=0.95\linewidth]{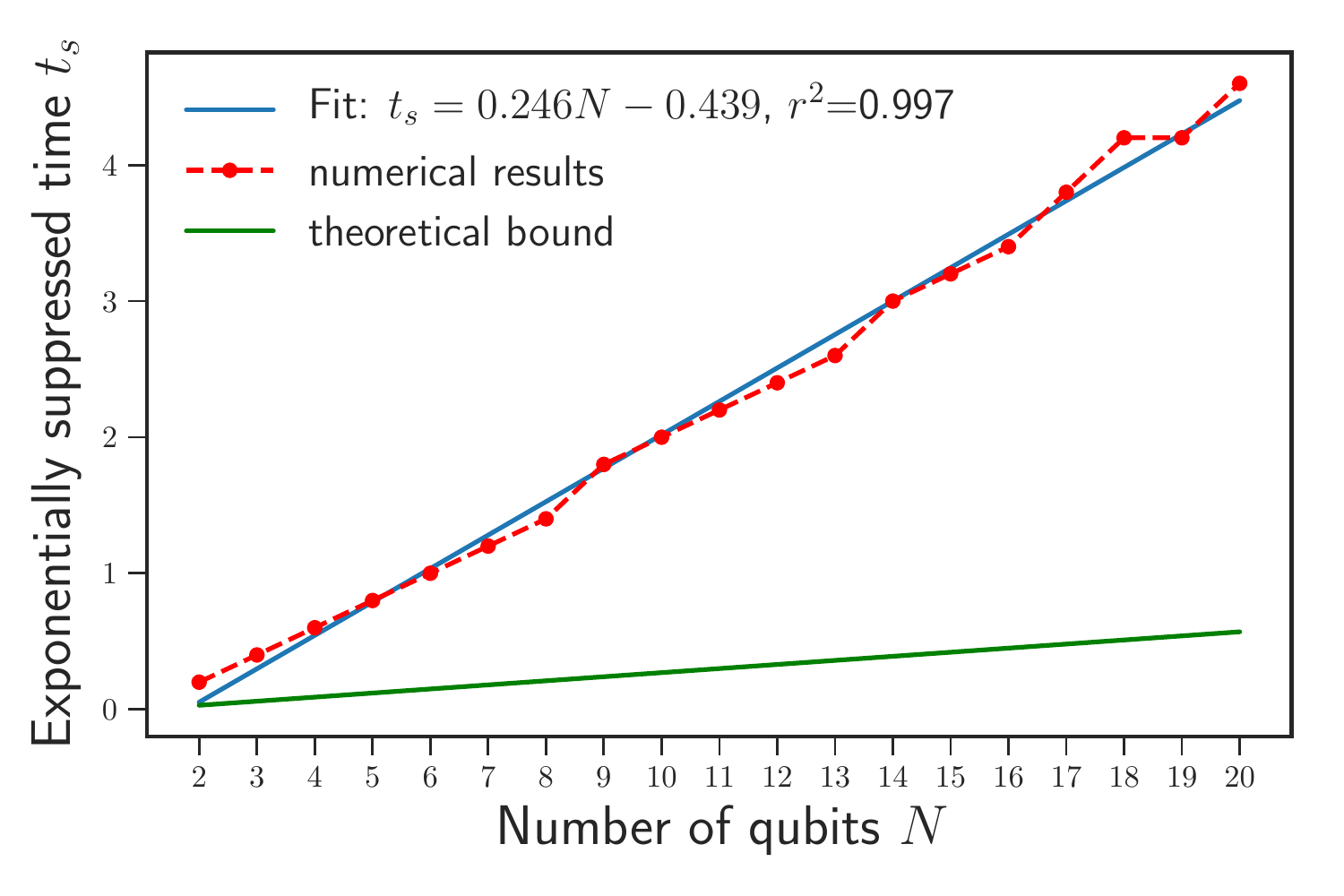}
\caption{Exponentially suppressed time $t_f$ (achieved fidelity $F<0.01$) versus the number of qubits ($N=2\to20$). The Lieb-Robinson bound gives a bound approximately of $t_f\sim 0.03N+c$. The exponentially suppressed time is defined as the time needed to make fidelity higher than 0.01. %{\color{red} Reason: constrained subspace.}
}
\label{tfes}
\end{figure}

The Lieb-Robinson bound gives a prediction about the size of the exponentially suppression region: $t_s\sim N/(6eJ)=0.03N$. Practically, we define the exponentially suppressed time as the time needed to make fidelity higher than 0.01. To see if it agrees with our numerical result, we plot the exponentially suppressed time as a function of the number of qubits in \cref{tfes}. Our numerical result is $t_s\sim 0.246N$ with a coefficient of determination $r^2=0.997$. We remark that the discrepancy between $0.246N$ and $0.03N$ is because our QAOAs only operate in the span of zero and single excitation subspaces, while the Lieb-Robinson  bound considered the full $N$-qubit Hilbert space.

\begin{comment}
    
Physically, this region can be treated using short-time perturbation theory by expanding the time evolution operators to low order. In a long time, this perturbation expansion becomes a poor approximation, which suggests that the dynamics in a long time is not a result of the direct qubit-qubit interaction. Instead, those propagation are due to the repeated propagation of information through intermediate qubits. A naive derivation gives
\begin{equation}
   \langle \sigma^z_i(t)\sigma^z_j\rangle_c = (J_{ij}t)^2+O(t^3).
\end{equation}
\end{comment}

\section{Summary}\label{SummarySection}
 We study the QAOA's success probability scaling as a function of circuit depth and the physical runtime  for implementing state transfer problems. By carefully utilizing the spectral properties of the QAOA Hamiltonians,  we obtain  analytic expressions for the success probability scaling as a function of the  circuit depth. At the low-circuit-depth and short-physical-duration limit, our analytic results reproduce the Grover-like quadratic speedup.  We further study the success probability scaling in numerically optimized QAOAs for a chain of up to $N=20$ qubits (limited by computational resources). These numerical experiments confirm the quadratic speed up and match with the Lieb-Robinson analysis of quantum speed limit, i.e., when the circuit depth $p$ is sufficiently large, with the increase of $t_f$, there are three different scenarios of success probability scaling: (1) exponentially suppressed region, (2) exponentially growing region and (3) steadily growing region. Treating QAOA optimization as a quantum control problem, we demonstrate the relation between the circuit depth and the controllability of QAOA: when the circuit depth is too low for a fixed distance state transfer, the control landscape possesses many locally optimal solutions that are not globally optimal and the QAOA becomes uncontrollable. Although the state transfer problem we considered here is relatively simple, %Notwithstanding a relatively simple problem state transfer is
 ,  our results offer valuable insights into the performance of  QAOAs by connecting its optimality to the Grover speedup and its success probability dependence on circuit-depth to the controllability  the QAOA ansatz.  To fully explore the application of QAOA,  however, more work remains to be done to study the effect of  realistic quantum noise on  QAOA implementations.

\clearpage
\appendix
\pagebreak
\widetext
\label{appendix}

\section{Optimal control solution and Pontryagin's maximum principle}\label{optimal-control-pontryagins-principle}

%Below, we first use optimal control theory and provide the optimal bang-bang solution for QAOA by solving the state transfer problem described in the last section.

In this appendix, we first solve the dynamical equation of our system in the span of zero and single excitation subspaces. Then we apply optimal control theory~\cite{stengel2012optimal,krotov1995global} to help design our state transfer protocol.

The dynamic of the system is governed by the Schr\"{o}dinger's equation (with \(\hbar=1\)): 
\begin{equation}
\frac{d|\psi(t)\rangle}{dt} = -i\hat{H}(t)|\psi(t)\rangle.
\label{Schrodinger}
\end{equation} 
Let \(|\psi(t)\rangle=\sum_{i=1}^N C_i(t)|\overline{i}\rangle\), where \(C_i(t)\)s are the complex amplitudes of the wave function defined in computational basis. We choose \(\vec{c}(t) =\{C_1(t),C_2(t),\cdots, C_N(t)\}\) as the dynamic variables for our problem. Substituting \(|\psi(t)\rangle=\sum_{i=1}^N C_i(t)|\overline{i}\rangle\) into Eq.~\eqref{Schrodinger}, we have
\begin{equation}
\begin{aligned}
\sum_{j=1}^N\frac{d}{dt}C_j(t)|\overline{j}\rangle &= -i\left\{s(t)\hat{H}_c+[1-s(t)]\hat{H}_B\right\}\sum_{j=1}^N C_j(t)|\overline{j}\rangle\\
& = -i\left\{s(t) |\overline{N}\rangle\langle \overline{N} |+[1-s(t)] \sum_{h=1}^{N-1} [\sigma_h^x\sigma_{h+1}^x+\sigma_{h}^y\sigma_{h+1}^y]\right\} \sum_{j=1}^N C_j(t)|\overline{j}\rangle\\
& = -is(t) C_N(t)|\overline{N}\rangle+\left\{-i[1-s(t)] \sum_{h=1}^{N-1} \left[\sigma_h^x\sigma_{h+1}^x+\sigma_{h}^y\sigma_{h+1}^y\right]\right\} \sum_{j=1}^N C_j(t)|\overline{j}\rangle .
\end{aligned}
\end{equation} 
Left multiplying both sides by $\langle \overline{j}|$, we get
\begin{equation}
\begin{aligned}
\frac{d}{dt}C_j(t)\langle \overline{j}|\overline{j}\rangle &= -is(t) C_N\langle j|\overline{N}\rangle+\left\{-i[1-s(t)]\langle \overline{j}| \sum_{h=1}^{N-1} [\sigma_h^x\sigma_{h+1}^x+\sigma_{h}^y\sigma_{h+1}^y]\right\} \sum_{k=1}^N C_k(t)|\overline{k}\rangle \\
&=-i\left\{s(t) C_N\langle \overline{j}|\overline{N}\rangle+[1-s(t)] \sum_{k=1}^N B_{j,k} C_k\right\},
\end{aligned}
\end{equation} 
where 
\begin{equation}
\begin{aligned}
B_{j,k} & = \langle \overline{j}|\sum_{h=1}^{N-1} [\sigma_h^x\sigma_{h+1}^x+\sigma_{h}^y\sigma_{h+1}^y]  |\overline{k}\rangle\\
& =   2\delta(k-j-1).
\end{aligned}
\end{equation}
Then we arrive in the dynamics equation in the form \(\dot{\vec{c}}(t) = f(\vec{c}(t),s(t))\).
\begin{equation}
\begin{aligned}
\frac{d}{dt}C_j(t) &=-i\left\{s(t) C_N\langle \overline{j}|\overline{N}\rangle+[1-s(t)] \sum_{k=1}^N B_{j,k} C_k(t)\right\}\\
& =-i\left\{s(t) C_N\delta_{jN}+2[1-s(t)] \sum_{k=1}^N    \delta(k-j-1)  C_k\right\}\\
& = \sum_{k=1}^{N} A_{j,k}C_k,
\end{aligned}
\end{equation} where 
\begin{equation}
A_{j,k} = -i\left\{s(t) \delta_{jN}+2[1-s(t)] \delta(k-j-1) \right \}.
\end{equation}

The cost function (action) for our state transfer problem is given by
\begin{equation}
J[\vec{c}(t_f)] = - |C_N(t_f)|^2,
\end{equation}
which only depends on the final state. Thus the problem we are solving is of Mayer type~\cite{stengel2012optimal}. Then the control Hamiltonian is linearly dependent to the conjugate momentum \(\vec{p}\) and control dynamics is 
\begin{equation}
H_{\text{control}} = \vec{p}^T\cdot f(\vec{c}(t),s(t))=\vec{p}^T\cdot\textbf{A}\cdot \vec{c},
\end{equation} 
such that the conjugate momentum is determined by the control Hamiltonian in the same way as that in the classical mechanics: 
\begin{equation}
\dot{\vec{p}} = -\partial_{\vec{c}} H_{\text{control}} (\vec{c},\vec{p},s).
\end{equation} 
With the fixed total time \(t_f\), the boundary condition for conjugate variable is given by 
\begin{equation}
\vec{p}(t_f) = \left.~\frac{\partial J(t)}{\partial \vec{c}}\right|_{t=t_f}.
\end{equation} 
Denote the components of \(\vec{p}\) to be $P_i(t)$. $P_i(t)$ should satisfy:
\begin{equation}
\begin{aligned}
\frac{d}{dt}P_j(t) &=-is(t) P_N\delta_{jN}+2[1-s(t)]P_{j+1}.
%& = -i\left\{s(t) P_N\delta_{jN}+2[1-s(t)]P_{j+1}\right\}^T.
\end{aligned}
\end{equation} 
Subsequently, the necessary and sufficient conditions for an optimal control \(s(t)\) is determined by: 
\begin{equation}
\frac{\partial \hat{H}}{\partial s} = 0,\frac{\partial^2 \hat{H}}{\partial s^2} \geq 0.
\end{equation} 
However, this cannot be applied to linear control problem where $\frac{\partial \hat{H}}{\partial s}$ is not a function of $s$. Pontryagin's principle comes to rescue, which replace two criteria with three new ones that are necessary and sufficient.

\begin{align}
    &H_{\text{control}} (\vec{c}^*,\vec{p}^*,s^*)\leq H_{\text{control}} (\vec{c}^*,\vec{p}^*,s), \forall t\in [0,t_f]\\
    & \vec{p}(t_f)=\partial_{\vec{c}}J[\vec{c}(t_f)],\\
    & \partial_t J[\vec{c}]+H_{\text{control}}(\vec{c}^*,\vec{p}^*,s^*)|_{t=t_f}=0
\end{align}

Since the control Hamiltonian is a linear function of the control parameter $s(t)(0\leq s(t)\leq 1)$ , $s(t)$ should be maximized when $\partial_s H_{\text{control}}<0$ and should be minimized when $\partial_s H_{\text{control}}>0$. The optimal control for QAOA is therefore determined from Pontryagin's theorem as follows: 
\begin{equation}
s(t)=\begin{cases}
0& \frac{\partial \hat{H}_{\text{control}}}{\partial s}>0\\
1 & \frac{\partial \hat{H}_{\text{control}}}{\partial s}<0
\end{cases}
\end{equation} 
Then, the best control solution \(s(t)\) is of bang-bang form, which corresponds to switching between two constant controls for each time duration. The bang-bang form of control contains abrupt switch between two values of $s(t)$ at time $t_0$ determined by $\partial_s H_{\text{control}}|_{t=t_0}=0$, or specifically as
\begin{equation}
  \vec{p}^T \cdot \mathbf{F}\cdot \vec{c}=0,
    \label{optimaleq}
\end{equation}
where the matrix elements of $F$ are given by
\begin{equation}
  F_{ij}= \delta_{iN}-2\delta(j-i-1).
\end{equation}
It is therefore trivial to verify whether a given control $s(t)$ as a function of $t$ is optimal or not. However, finding the `optimal' control is generally hard due to the mutual dependency of control and system dynamics. A brute force search on switching time is already computationally formidable but is still unable to find the optimal control without specifying the number of bangs. 

\clearpage

\section{Details of our numerical optimization}

In this appendix, we provide more details on our numerical optimizations.  \Cref{table-tfp} summarizes the parameters $t_f$ and $p$ we performed grid search on. In the first round of grid search (Run 1), we investigate  the general performance of optimized QAOAs with different fixed total run time $t_f$ and circuit depth $p$. To investigate the perfomance of optimized QAOA within or near the exponentially suppressed region, we performed the second round of grid search which scanned more densely spaced total run time over a smaller interval. 
\cref{summary-n-r} overviews our numerical results and their implications. 

In these tables, we adopt the MATLAB style to represent arrays, for example, a "0.2:0.2:1.6" corresponds to an array of "[0.2 0.4 0.6 0.8 1.0 1.2 1.4 1.6] and a "1:5" corresponds an array of "[1 2 3 4 5]".

\begin{table}[h]
\centering
\caption{Parameters of the first and the second round of grid searches over $t_f$ and $p$. The first round were used to determine the general performance of optimized QAOAs. The second round were used to investigate the performance of optimized QAOAs near or within the exponentially suppressed region that we have identified.}
\begin{tabular}{|l|l|l|l|l|l|l|l|l|l|l|l|}
\hline
Runs                                                                          & $N$     & 2           & 3           & 4           & 5           & 6           & 7           & 8           & 9           & 10          & 11          \\ \hline
\multirow{2}{*}{\begin{tabular}[c]{@{}l@{}}Run 1: \\ ~\end{tabular}}   & $p$   & 1:7         & 1:8         & 1:9         & 1:10        & 1:11        & 1:12        & 1:13        & 1:14        & 1:15        & 1:16        \\ \cline{2-12} 
                                                                              & $t_f$ & 1:8         & 1:10        & 1:12        & 1:14        & 1:16        & 1:18        & 1:20        & 1:22        & 1:25        & 1:27        \\ \hline
\multirow{2}{*}{\begin{tabular}[c]{@{}l@{}}Run 2:\\ ~\end{tabular}} & $p$   & 1:7         & 1:8         & 1:9         & 1:10        & 1:11        & 1:12        & 1:13        & 1:14        & 1:15        & 1:16        \\ \cline{2-12} 
                                                                              & $t_f$ & 0.2:0.2:1.6 & 0.2:0.2:2.0 & 0.2:0.2:2.4 & 0.2:0.2:2.8 & 0.2:0.2:3.2 & 0.2:0.2:3.6 & 0.2:0.2:4.0 & 0.2:0.2:4.4 & 0.2:0.2:5.0 & 0.2:0.2:5.4 \\ \hline
                                                                              & $N$      & 12          & 13          & 14          & 15          & 16          & 17          & 18          & 19          & 20          &             \\ \hline
\multirow{2}{*}{\begin{tabular}[c]{@{}l@{}}Run 1: \\ ~\end{tabular}}   & $p$   & 1:17        & 1:18        & 1:19        & 1:20        & 1:2:23      & 1:2:25      & 1:2:27      & 1:2:29      & 1:2:31      &             \\ \cline{2-12} 
                                                                              & $t_f$ & 1:30        & 1:32        & 1:34        & 1:36        & 1:2:39      & 1:2:41      & 1:2:43      & 1:2:45      & 1:2:47      &             \\ \hline
\multirow{2}{*}{\begin{tabular}[c]{@{}l@{}}Run 2:\\ ~\end{tabular}} & $p$   & 1:17        & 1:18        & 1:19        & 1:20        & 1:2:23      & 1:2:25      & 1:2:27      & 1:2:29      & 1:2:31      &             \\ \cline{2-12} 
                                                                              & $t_f$ & 0.2:0.2:6.0 & 0.2:0.2:6.4 & 0.2:0.2:6.8 & 0.2:0.2:7.2 & 0.2:0.4:7.8 & 0.2:0.4:8.2 & 0.2:0.4:8.6 & 0.2:0.4:9.0 & 0.2:0.4:9.4 &             \\ \hline
\end{tabular}
\label{table-tfp}
\end{table}

\begin{table}[h]
\centering
\caption{Parameters of the grid search over $p$ with no constraints on $t_f$. The obtained numerical results are presented and analyzed in \Cref{No-limit-tf}.}
\begin{tabular}{|l|l|l|l|l|l|l|l|l|l|l|l|l|l|l|l|l|l|l|l|l}
\hline
$N$  & 2 & 3 & 4 & 5 & 6 & 7 & 8 & 9           & 10 & 11 & 12 & 13  & 14  & 15  & 16   & 17 & 18          & 19  & 20\\ 
\hline
$p$  & 1:7 & 1:8 & 1:9 & 1:10 & 1:11 & 1:12 & 1:13 & 1:14 & 1:15 & 1:16  & 1:17 & 1:18 & 1:19 & 1:20  & 1:22 & 1:24 & 1:26 & 1:28 & 1:30 \\ 
\hline
\end{tabular}
\label{table-tfp}
\end{table}

\begin{table}[h]
\centering
\caption{Overview of our numerical results. We consider the performance of optimized QAOAs with unlimited $t_f$ in \Cref{No-limit-tf} and that with fixed $t_f$ in \Cref{numerical-results}. In \Cref{success-probability-p-versus-the-number-of-switches-p}, we study the maximum achievable fidelity $F$ as a function of circuit depth $p$ , and investigated the controllability of QAOAs. In \Cref{success-probability-p-versus-total-run-time-t_f}, we study the maximum achievable fidelity $F$ as a function of total run time $t_f$, and investigate the Lieb-Robinson type quantum speed limit emerged in QAOAs.}
\label{summary-n-r}
\begin{tabular}{l|l|l|l|l|l|l}
\hline
& Figures  & Max achi. $F$ & Physical run time $t_f$ & Circuit depth $p$        & Number of qubits $N$ & Goal \\ \hline
\multirow{3}*{\Cref{No-limit-tf}} & \Cref{fit2-20-quadratic}  & Max achi. $F$ & unlimited $t_f$ & (a): $p=$[1:6]; (b): $p=$[1:11]   & $N=2\to20$ & QAOA \\ 
~ & \Cref{N10}  & N/A & unlimited $t_f$ & $p=15$  & $N$=10 & bang-bang sol. \\ 
unlimited $t_f$ & \Cref{landscape}  & $z$ variable & unlimited $t_f$ & $p=2,4$ & $N=3$ & Landscape \\ 
\hline
\multirow{4}*{\Cref{success-probability-p-versus-the-number-of-switches-p} }  & \Cref{p6-13}(a) & $y$ variable  & $t_f=6$ & $x$ variable: [1:9]  & [5,10,15,19] &  Controllability    \\ 
~ & \Cref{p6-13}(b) & $y$ variable  & $t_f=13$ & $x$ variable: [1:9] & [5,10,15,19] &      \\
~ & \Cref{p2-16}(a) & $y$ variable  & sufficiently large & $x$ variable: [1:6]  & [2,4,6,8] &      \\
Fixed $t_f$  & \Cref{p2-16}(b) & $y$ variable  & sufficiently large & $x$ variable: [1:15] & [10,12,14,16] &      \\ \hline
%~ & \Cref{p10-15}(a) & $y$ variable  & [1:3:22] & $x$ variable: [1:8]  & $N=10$ &      \\
%~& \Cref{p10-15}(b) & $y$ variable  & [1:4:35] & $x$ variable: [1:12] & $N=15$ &      \\ \hline
\multirow{8}*{\Cref{success-probability-p-versus-total-run-time-t_f}}    & \Cref{p2-9-}(a) & $y$ variable  & $x$ variable: all &  $p=2$ & [5,10,15,19] &   LR-bound   \\
~ & \Cref{p2-9-}(b) & $y$ variable  & $x$ variable: all  & $p=9$ & [5,10,15,19] &      \\
& \Cref{2-16-m1}(a) & $y$ variable  & $x$ variable: all &  sufficiently large  & [2,4,6,8] &      \\
~& \Cref{2-16-m1}(b) & $y$ variable  & $x$ variable: all &  sufficiently large  & [10,12,14,16] &      \\
Fixed $p$ & \Cref{10-15-F-tf}(a) & $y$ variable  & $x$ variable: [1:25]  &  sufficiently large  & $N=10$ &      \\
~& \Cref{10-15-F-tf}(b) & $y$ variable  & $x$ variable: [1:30]& sufficiently large  & $N=15$ &      \\ 
~& \Cref{tf-N-2-19} & $F>0.99$  &   $y$ variable     &   sufficiently large     & $x$ variable: [2:20]      &      \\
%\multirow{4}*{\Cref{small-tf}}    & \Cref{p27}(a) & $y$ variable  & $x$ variable: all &  $p=2$ & [4:2:20] & LR-bound     \\
%~ & \Cref{p27}(b) & $y$ variable  & $x$ variable: all  & $p=7$ & [4:2:20] &      \\
%Fixed $p$ & \Cref{p46-20}(a) & $y$ variable  & $x$ variable: all &  fully controllable  & [4:2:20] &      \\
~& \Cref{tfes} & $F<0.01$  &   $y$ variable     &   sufficiently large     & $x$ variable: [2:20]      &      \\ \hline

\end{tabular}
\end{table}

% \subsection{Projected gradient descent}\label{projected-gradient-descent}

% The variables are \((t_1,t_2,t_2' \cdots,t_{p},t_{p}')\) and the fixed total time gives: 
% \begin{equation}
% t_1+\sum_{i=2}^{p}(t_i +t_i')= t_f,~t_i,t_i'\geq 0.
% \end{equation} 
% which is convex and an additional constraint on variables. We use projection gradient descent method to optimize success probability over possible time of switches. For general minimization problem over a convex set $C$, the projected gradient descent (PGD) algorithm iteratively generates the sequence {$x_k$} via
% \begin{eqnarray}
% x_{k+1} &=& \pi_{C}(x_k-\alpha_{k} \nabla,f(x_{k}))
% \end{eqnarray}
% with $\alpha_{k}>0$ is the step-size determined by optimizer and $\pi_{C}$ is the projection operator onto the convex set. 

\end{document}